\documentclass[fleqn,usenatbib]{mnras}

\usepackage{newtxtext,newtxmath}

\usepackage[T1]{fontenc}

\DeclareRobustCommand{\VAN}[3]{#2}
\let\VANthebibliography\thebibliography
\def\thebibliography{\DeclareRobustCommand{\VAN}[3]{##3}\VANthebibliography}

\usepackage{graphicx}	
\usepackage{amsmath}	
\usepackage{float}
\usepackage{multirow}

\newcommand\msun{$M_\odot$} 
\newcommand\teff{$T_\mathrm{eff}$} 
\newcommand\logg{log~\textit{g}} 
\newcommand\kep{\textit{Kepler}} 
\newcommand\cn{[C/N]} 
\newcommand\feh{[Fe/H]}
\newcommand\cfe{[C/Fe]}
\newcommand\nfe{[N/Fe]}
\newcommand\mgh{[Mg/H]}
\newcommand\mgfe{[Mg/Fe]}

\newcommand{\changed}[1]{#1}
\graphicspath{{./}{figures/}}

\title[Nature vs. Nurture]{Nature vs. Nurture: Distinguishing Effects from Stellar Processing and Chemical Evolution on Carbon and Nitrogen in Red Giant Stars}

\author[J. Roberts et al.]{%
John D. Roberts$^{1,2}$\thanks{E-mail: roberts.2158@osu.edu}
Marc H.~Pinsonneault$^{1,2}$
Jennifer A. Johnson$^{1,2}$
Joel C. Zinn,$^{3,4}$
David H. Weinberg,$^{1,2}$
\newauthor
Mathieu Vrard$^{1,5}$,
Jamie Tayar$^{6}$, 
Dennis Stello,$^{7,8,9}$ 
Beno\^it Mosser,$^{10}$ 
James W.\  Johnson,$^{1,2,11}$
\newauthor
Kaili Cao,$^{2,12}$ 
Keivan G.\ Stassun,$^{13}$  
Guy S.\ Stringfellow,$^{14}$  
Aldo Serenelli,$^{15,16}$
Savita Mathur,$^{17,18}$ 
\newauthor
Saskia Hekker,$^{19,20,21}$
Rafael A. García,$^{22}$
Yvonne P. Elsworth,$^{23}$
Enrico Corsaro$^{24}$ 
 \\
$^{1}$Department of Astronomy, The Ohio State University, 140 W. 18th Ave., Columbus, OH 43210, USA\\
$^{2}$Center for Cosmology and Astroparticle Physics, The Ohio State University, Street Address, Columbus, OH 43210, USA\\
$^{3}$Department of Astrophysics, American Museum of Natural History, Central Park West at 79th Street, New York, NY 10024, USA\\
$^{4}$NSF Astronomy and Astrophysics Postdoctoral Fellow.\\
$^{5}$Laboratoire Lagrange, Universit\'e C\^ote d’Azur, Observatoire de la C\^ote d’Azur, CNRS, Parc
Valrose, 06108 Nice, France \\
$^{6}$Department of Astronomy, University of Florida, Bryant Space Science Center, Stadium Road, Gainesville, FL 32611, US\\
$^{7}$School of Physics, University of New South Wales, Sydney, NSW 2052, Australia; \\
$^{8}$Sydney Institute for Astronomy, School of Physics, A28, The University of Sydney, NSW 2006, Australia\\
$^{9}$ARC Centre of Excellence for All Sky Astrophysics in 3 Dimensions (ASTRO 3D), Australia\\
$^{10}$LESIA, Observatoire de Paris, Universit\'e PSL, CNRS, Sorbonne Universit\'e, Universit\'e Paris-Cit\'e, 92195 Meudon, France; \\
$^{11}$The Observatories of the Carnegie Institution for Science, 813 Santa Barbara St,. Pasadena, CA 91101, USA\\
$^{12}$Department of Physics, The Ohio State University, 191 W. Woodruff Ave., Columbus, OH 43210, USAA\\
$^{13}$Department of Physics and Astronomy, Vanderbilt University, Nashville, TN 37235, USA \\
$^{14}$Center for Astrophysics and Space Astronomy, Department of Astrophysical and Planetary Sciences, University of Colorado, 389 UCB, Boulder, CO 80309-0389, USA
$^{15}$Institute of Space Sciences (ICE, CSIC), Carrer de Can Magrans S/N, 08193 - Cerdanyola del Valles, Spain \\
$^{16}$Institut d'Estudis Espacials de Catalunya (IEEC), Carrer Gran Capità, 2-4, 08034 - Barcelona, Spain \\
$^{17}$Instituto de Astrof\'isica de Canarias (IAC), E-38205 La Laguna, Tenerife, Spain\\
$^{18}$Universidad de La Laguna (ULL), Departamento de Astrof\'isica, E-38206 La Laguna, Tenerife, Spain \\
$^{19}$Heidelberg University, Centre for Astronomy, Landessternwarte,
K\"onigstuhl 12, Heidelberg, Germany\\
$^{20}$Heidelberg Institute for Theoretical Studies (HITS), Schloss-Wolfsbrunnen Weg 35, Heidelberg, Germany\\
$^{21}$Stellar Astrophysics Centre, Aarhus, Denmark\\
$^{22}$Universit\'e Paris-Saclay, Universit\'e Paris Cit\'e, CEA, CNRS, AIM, 91191, Gif-sur-Yvette, France \\
$^{23}$School of Physics and Astronomy, University of Birmingham, Edgbaston, Birmingham B15 2TT, UK \\
$^{24}$INAF - Osservatorio Astrofisico di Catania, Via S. Sofia 78, I-95123, Italy 
}

\date{Accepted XXX. Received YYY; in original form ZZZ}

\pubyear{2023}

\begin{document}
\label{firstpage}
\pagerange{\pageref{firstpage}--\pageref{lastpage}}
\maketitle

\begin{abstract}
The surface \cn\ ratios of evolved giants are strongly affected by the first dredge-up (FDU) of nuclear-processed material from stellar cores. C and N also have distinct nucleosynthetic origins and serve as diagnostics of mixing and mass loss. We use subgiants to find strong trends in the birth [C/N] with [Fe/H], which differ between the low-$\alpha$ and high-$\alpha$ populations. We demonstrate that these birth trends have a strong impact on the surface abundances after the FDU. This effect is neglected in current stellar models, which used solar-scaled C and N. We map out the FDU as a function of evolutionary state, mass, and composition using a large and precisely measured asteroseismic dataset in first-ascent red giant branch (RGB) and core He-burning, or red clump (RC), stars. We describe the domains where [C/N] is a useful mass diagnostic and find that the RC complements the RGB and extends the range of validity to higher mass. We find evidence for extra mixing on the RGB below \feh$=-0.4$, matching literature results, for high-$\alpha$ giants, but there is no clear evidence of mixing in the low-$\alpha$ giants. The predicted signal of mass loss is weak and difficult to detect in our sample. We discuss implications for stellar physics and stellar population applications.
\end{abstract}




\section{Introduction} \label{sec:intro}

The surface carbon and nitrogen abundances of low-mass stars are powerful diagnostics of both stellar evolution and galactic chemical evolution. During the main-sequence and subgiant phases, a star's surface abundances reveal its birth composition, which is the result of contributions from massive star winds, core-collapse supernovae, and asymptotic giant branch stars. As a star expands onto the RGB, its convective envelope dips deeper into the interior of the star, bringing the results of H-burning to its surface, altering these abundances \citep{Iben_1967}. Notably, the CNO cycle severely depletes carbon and enriches nitrogen, so mixing this processed material to the surface results in sharp drops in the C$^{12}$/C$^{13}$ and \cn\ ratios. The ultimate depth of the surface convection zone, and thus the degree of these drops, depends on the mass and metallicity of the star \citep{Sweigart_1978}. The mass-dependence of first dredge-up (FDU) has been exploited to derive ages for field red giants, producing the most extensive maps of galactic chronology \citep[eg.][]{Ness_2015, Martig_thickdisk}. However, large-scale applications of these methods run the risk of introducing errors caused by complicating physical effects. While standard stellar theory predicts no further changes to surface abundances on the RGB, observations have shown consistently lower \cn\ and carbon isotopic ratios for upper RGB stars versus the lower RGB,  implying the existence of ``extra mixing'' \citep[eg.][]{Gilroy_1989, Gratton_2000}. Finally, if there is mass loss at the tip of the RGB, then stars in the core-helium-burning red clump phase will have lower masses than predicted from their \cn, which was set by the FDU. 

The initial \cn\ in stars is expected to change over galactic history as a result of chemical evolution. Carbon is produced from He-burning in massive stars and low-mass AGB stars, while nitrogen is a secondary product from H-burning in intermediate-mass (4-7 \msun) AGB stars \citep{Timmes_1995, Henry_2000, Kobayashi_2020}. Neither of these elements are primarily produced in Type Ia supernovae, which are efficient sources of iron. Because these elements come from different sources, there is no reason to expect carbon and nitrogen to scale directly with iron. Indeed, the contributions to C from Type II supernovae could yield a correlation with the $\alpha$ elements, which are mainly produced in Type II supernovae, rather than with Fe, which is produced in both Type II and Type Ia. The solar neighbourhood shows two sequences in the [Mg/Fe]-\feh\ plane \citep[e.g.][]{Fuhrmann1998, Bensby_2003}. These two populations reflect different contributions from prompt enrichment, namely Type II supernovae, compared to a delayed component, such as AGB stars and Type Ia supernovae. Investigating the \cn\ trends as both a function of metallicity and of [$\alpha$/Fe] is critical for establishing the initial \cn\ based on elements that are not changed during FDU and subsequent evolution.

Subgiant stars can provide an accurate picture of the birth \cn. Standard stellar theory, computed using 1-D models, predicts limited mixing in stars before the first dredge-up, indicating the pre-FDU surface \cn\ should follow the birth abundances. Rotational mixing is not expected to have a noticeable effect on surface abundances pre-RGB for lower-mass ($< 3$\msun) stars \citep{Pinsonneault_1989}. The effects of gravitational settling, where heavier isotopes drop out of the bottom of the convective envelope and therefore are removed from mean surface abundances, may appear for F and G stars. However, since these materials are deposited right below the convective zone, they are quickly swept back up as the star expands on the subgiant branch \citep[eg.][]{Souto_2019}. Because of this, these pre-FDU subgiants provide an avenue to probe the changes and trends in the birth mixture of stars.

The FDU marks an irreversible change in the surface \cn. More massive stars develop deeper convective zones and hotter internal temperatures, so their \cn\ drops more significantly. The dredge-up is not purely dependent on mass, however. The composition of a star affects its internal opacities and therefore its energy transport. Higher metallicity stars have shallower convective zones \citep{Karakas_2014}, and this can impact the magnitude of the \cn\ drop observed during FDU. Further, the impact of pre-FDU surface abundances on post-FDU abundances is almost universally ignored, and it is not readily apparent how much of an effect pre-FDU abundances should have. Since the fraction of processed material in the envelope after the FDU is unknown, it is also unknown if the original \cn\ abundances get diluted to negligible levels, or remain relevant after dredge-up.

Contrary to standard model predictions, there is clear evidence for in situ changes in \cn\ for some stellar populations. Extensive studies of isotopic and elemental abundances for C, N, and O have shown that FDU is insufficient to explain the observations in metal-poor field stars and globular cluster stars \citep[e.g.,][]{Kraft_1994, Charbonnel_1995, Gratton_2000, Shetrone_2003, Takeda_2019}. \changed{Additionally, drops in the lithium abundance and carbon isotope ratio have been observed in open clusters across larger metallicity ranges \citep[e.g.][]{Gilroy_1989, Drazdauskas_2016, McCormick_2023}.} The mechanism of this ``extra mixing'' has yet to be determined. One frequently discussed mechanism is thermohaline mixing. Thermohaline mixing, or ``salt-finger instability,'' is a diffusive mixing process that occurs in regions with an inverted mean molecular weight gradient. This allows the material to mix even in regions stable against convection. While some works have shown thermohaline mixing to be a viable candidate \citep[e.g.,][]{Charbonnel&Zahn_2007}, other works find that thermohaline mixing alone does not fully explain observed abundance behaviors \citep[e.g.,][]{Denissenkov_2010, Traxler_2011, Tayar&Joyce_2022}. Additionally, it is known that this extra mixing is both metallicity-dependent and mass-dependent \citep{Charbonnel&Lagarde_2010, Shetrone_2019, Magrini_2021}. Examining \cn\ in such a way that accounts for extra mixing and provides additional constraints is therefore imperative.

Properly calibrated, an empirical \cn-mass-evolutionary state relation can be used as a direct test of stellar theory. Models make firm predictions about the degree of dredge-up, the mass and composition dependence, and the location on the HR diagram where dredge-up occurs \citep[e.g.,][]{Salaris_2015}. By comparing theoretical predictions with our observed trends, we can determine if there are aspects affecting these processes not completely covered by current stellar models. To perform these tests, we need compositions, evolutionary states, and masses for a comprehensive sample of stars. Star clusters have been excellent laboratories for these investigations because their populations have one age and one initial composition. The stars in later stages of evolution are close in mass, because of the short length of time post-hydrogen burning. 

\cite{Salaris_2015} used the BaSTI stellar evolution database \citep{Basti_1, Basti_2} to calculate a purely theoretical calibration of ages. They found qualitative agreement with clusters of known ages but did not believe the method would be accurate enough for individual stars. \citet{Casali_2019} created an empirical relationship between \cn\ and age using clusters of known ages but also did not find a strong enough relation to warrant application to individual stars. However, the number, age range, and metallicity range of nearby clusters are very limited. 

Asteroseismology provides a method to infer stellar mass through the characterization of stellar oscillations that cause brightness variations observed on the stellar surface \citep{Bedding2011, Mosser_2012, Vrard_2016}. 
Given samples of RGB and RC stars with known masses and chemical abundances, it is then possible to calibrate relationships that provide mass and age based on stellar chronometers that are more readily obtained. Large spectroscopic surveys are a natural resource. 

\cite{Martig_2016} used the APOKASC catalogue \citep{apokasc1} as a sample with known masses from seismic analysis and spectroscopic parameters to calibrate the relationship using individual stellar observations, a method similar to what will be employed here. They used ages from \cn\ to examine the radial age gradient in the thick disc \citep{Martig_thickdisk}. However, they only had about $\sim1500$ stars from the first APOKASC sample, fit in C/N, [(C+N)/M], [Fe/H] and mass, and their method resulted in systematic underestimates for higher-mass stars. \cite{Lagarde_2017} used STAREVOL model grids to look at \cn\ versus mass for RGB and RC stars and to examine the strength of extra mixing effects. None of the above studies focused on the effects of the birth mixture on post-FDU \cn\ values, though \cite{Martig_2016} did include a [(C+N)/M] axis as a means of allowing the fit to account for different birth values of carbon and nitrogen, since [(C+N)/M] is not expected to change during dredge-up. 

In this paper, we take advantage of the new APOKASC3 catalog (Pinsonneault et al., in prep), which features more precisely characterized samples of stars with measured abundances and seismic parameters than previously available to obtain improved fits for the relationship between mass, metallicity and \cn. We separate the stars based on evolutionary state and [$\alpha$/Fe]. We examine the effect of the dredge-up directly by comparing pre-dredge-up \cn\ values for subgiants to post-dredge-up for RGB and RC stars, as well as the strength of extra mixing effects by comparing different evolutionary states.

In Section \ref{sec:sample} we outline the source of our data and the criteria used to separate the different samples of stars. In Section \ref{sec:fitting} we present functions describing the \cn\ of the various samples and how they were obtained. In Section \ref{sec:analysis} we compare the \cn\ functions of different samples to explore how \cn\ changes across the post-main-sequence life of a star. In Section \ref{sec:discussion} we discuss the applicability and limitations of using these functions to estimate mass and compare our results with those of previous works. In Section \ref{sec:conclusion} we summarize our results and discuss avenues of future study.

\section{Data and Sample Selection} \label{sec:sample}

Our goal is to understand the birth mixture of stars, how the mixture is then modified by the first dredge-up, and the conditions under which extra mixing sets in on the giant branch. Fortunately, these effects can be distinguished by studying distinct populations across the HR diagram. Spectroscopy and asteroseismology are crucial tools in this regard; we therefore begin by describing our data sources and then proceed to sample selection.

\subsection{Data Sources}\label{data:sources}
\subsubsection{Spectroscopic Data from APOGEE}\label{data:spectro}

The Apache Point Observatory Galactic Evolution Experiment \citep[APOGEE;][]{Majewski2017} was part of the Sloan Digital Sky Survey (SDSS), in particular SDSS-III \citep{eisenstein2011} and SDSS-IV \citep{blanton2017}. It collected high-resolution H-band spectra using dual APOGEE spectrographs \citep{Wilson2019} at the 2.5-meter Sloan Foundation Telescope \citep{Gunn2006} at Apache Point Observatory and the 2.5-meter Ir{\'e}n{\'e}e DuPont telescope \citep{Bowen1973} at Las Campanas Observatory. In total, APOGEE observed over 650,000 stars in the Milky Way. Abundances, effective temperatures, and surface gravities used in this paper come from the 17th data release of SDSS \citep[][;DR17]{dr17}. The spectra were reduced by the APOGEE data reduction pipeline \citep{Nidever2015}. The stellar parameters and abundances were determined by the APOGEE Stellar Parameter and Chemical Abundances Pipeline \citep[ASPCAP;][]{aspcap}, which compares the observed spectra with a large grid of synthetic spectra along 7 axes: \teff, \logg, [M/H], [$\alpha$/M], [C/M], [N/M], and microturbulence. $\chi^2$ optimization is used to find the best-fit spectrum in those dimensions. Abundances of individual elements were measured around small areas of the spectra with absorption lines from that element. A $\chi^2$ optimization was done using synthetic spectra which varied the element of interest while the other axes remained fixed at the previously determined values. For example, to determine [Mg/M], synthetic spectra with different [$\alpha$/M] were compared to the observed spectrum around Mg lines. The effective temperatures, surface gravities, and abundances are then placed on an absolute scale in a post-processing calibration step \citep{jonsson2020}. Cases with suspect or bad overall fits are flagged. A description of the APOGEE flags can be found in \citet{jonsson2020}.
 
\subsubsection{Seismic Parameters from \kep}\label{data:APOKASC}

Masses and evolutionary states for giants come from the APOKASC catalog. APOKASC includes stars targeted by the \kep\ mission \citep{Kepler} and the APOGEE survey. There have been 3 catalogs in total \citep[][Pinsonneault et al., in prep]{apokasc1, apokasc2}, with a total of 15,779 evolved giants in the third complete data set. 10,004 of these stars constitute the ``gold sample'' of high-quality measurements that we adopt for our sample. Stellar parameters from seismology are calculated through 10 independent pipelines and compared to ensure accuracy, and the average of those pipelines is then added to the catalogue. A more complete description of the processes can be found in Pinsonneault et al. (in prep).

The two asteroseismic parameters of interest for our purposes are the mass and evolutionary state flag. Asteroseismology can be used to infer evolutionary states and to calibrate spectroscopic evolutionary state predictions. For a discussion on how these predictions are made, see \citet{Elsworth_2019, Warfield_2021}, and Pinsonneault et al. (in prep).

\subsection{Sample Selection}\label{data:samples}

Since our goal is to understand the birth abundances of stars and how the abundances change as the star evolves, we need the ability to distinguish between effects from chemical evolution and stellar processing. To this end, we define several different samples that will allow us to isolate these effects from each other. First, to separate the effects of the different chemical history channels of our galaxy, we divided the stars into low-$\alpha$ and high-$\alpha$ samples as detailed in Section \ref{data:alpha}. Second, to separate the effects of birth variation, FDU, and extra mixing, we isolate stars at different stages in their evolution. The states used in this paper and the sections in which they are defined are: pre-dredge-up subgiants (Section \ref{data:subgiants}), lower red giant branch stars (Section \ref{data:LowRGB}), upper red giant branch stars (Section \ref{data:HighRGB}), and red clump stars (Section \ref{data:clump}). 

Across all samples, we removed stars that have complicating phenomena that alter evolution, such as a close binary companion or known young stars. Any star flagged as a possible young cluster member, emission line star, MIR-detected candidate cluster member, part of the eclipsing binary program, or part of a W3/4/5 star-forming complex were removed because the ASPCAP stellar parameters become unreliable for these stars.

Additionally, stars at risk of having less reliable data were removed preemptively. If a star was missing a relevant parameter, namely \logg, \teff, \cfe, \nfe, \mgfe, \feh, and mass, and associated errors, it was removed entirely. Additionally, for stars where seismic data was used, any star with fewer than 3 quarters of \kep\ data or any star that had $\nu_\mathrm{max} < 1\;{\rm \mu Hz}$ or $\nu_\mathrm{max} < 220\;{\rm \mu Hz}$ was rejected as those measurements were not considered to be reliable enough to use. 

\begin{figure}
    \centering
    \includegraphics[width=.49\textwidth]{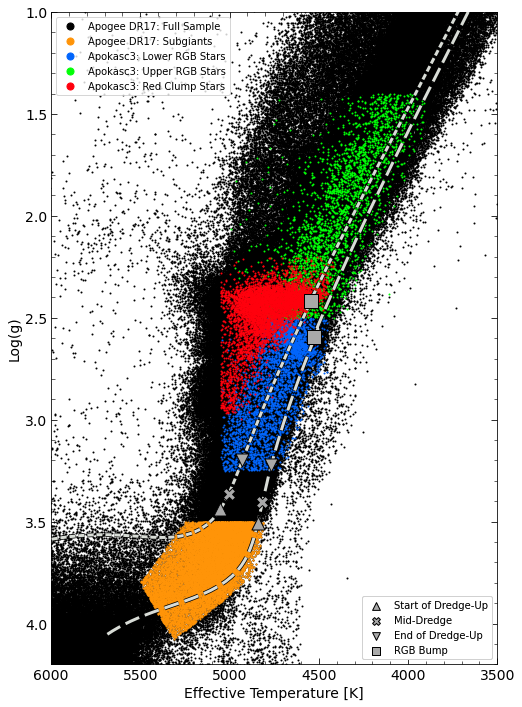}
    \caption{The samples in this paper with evolution tracks and the full DR17 samples for context. The lines are MIST stellar evolution tracks of 1 \msun\ (dotted) and 1.6 \msun\ (dashed) at solar metallicity. These tracks have markers denoting the onset, midpoint, and completion of the FDU as well as the beginning of re-contraction at the RGB bump. Due to the FDU beginning and ending slowly, the onset and completion are marked at the points where 10\% and 90\% of the total \cn\ change has occurred, respectively. Unlike the subgiants, not all giants from APOGEE are used because they are restricted to the APOKASC3 sample.}
    \label{fig:HR}
\end{figure}

\begin{figure}
    \centering
    \includegraphics[width=.45\textwidth]{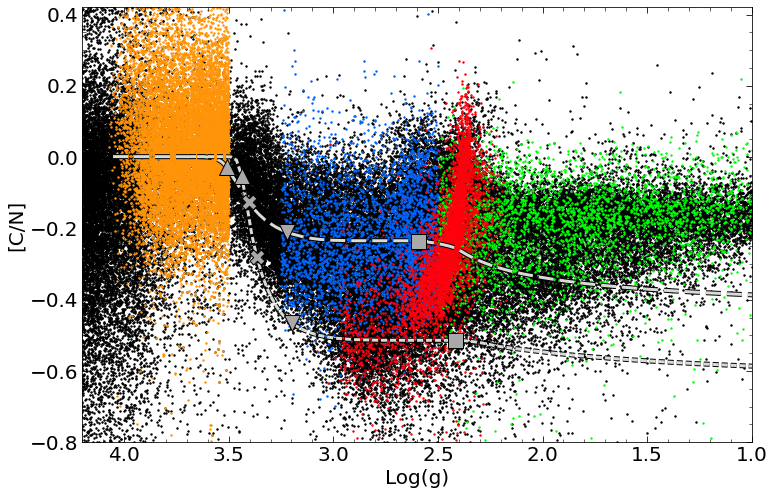}
    \caption{\cn\ versus \logg\ for the APOGEE sample, with the same tracks and colouring as Figure \ref{fig:HR}.}
    \label{fig:cnlogg}
\end{figure}

Figure \ref{fig:HR} shows the entire APOGEE DR17 sample in the \logg-\teff\ plane with the samples highlighted by different colours. We also show MIST evolutionary tracks \citep{MIST_0,MIST_1}, which were computed using MESA \citep{MESA_1,MESA_2,MESA_3,MESA_4}. The tracks are for solar metallicity for 1 and 1.6 \msun\ stars, a range that spans the majority of stars in our sample. Figure \ref{fig:cnlogg} shows the same samples and tracks, but in the \cn-\logg\ plane, where the FDU can be observed directly. 

\subsubsection{High-$\alpha$ and Low-$\alpha$ Populations}\label{data:alpha}

\begin{figure}
    \centering
    \includegraphics[width=.45\textwidth]{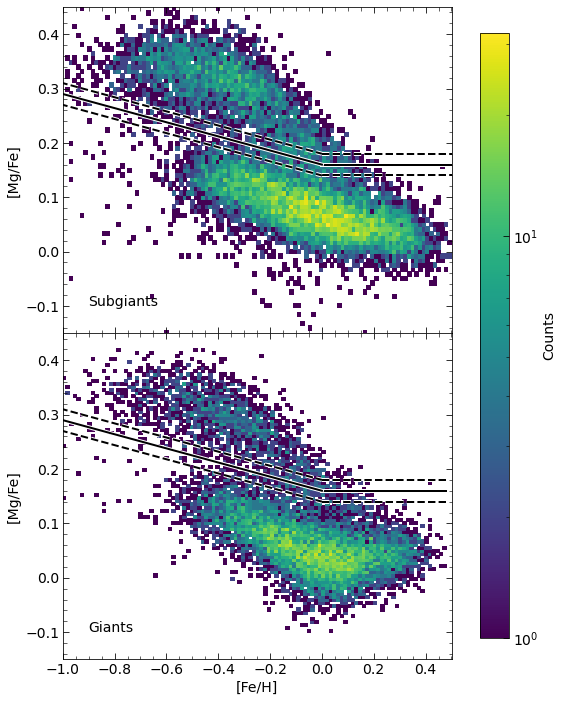}
    \caption{Density in the \mgfe-\feh\ plane for the APOGEE DR17 subgiant sample and the APOKASC3 giant sample. The solid line indicates the boundary between high- and low-$\alpha$ populations used in this paper, and the dashed lines show boundary shifted by 0.02 dex which reflects the actual cuts used to define the samples.}
    \label{fig:Alphapop}
\end{figure}

To separate the high-$\alpha$ from low-$\alpha$ stars, we adopt the criteria used by \cite{Weinberg_2019} with some small adjustments. Because there are systematic offsets in abundances between APOGEE data releases, we shift the separation criterion upwards by .04 dex to account for the differences between DR14 (used in \cite{Weinberg_2019}) and DR17 (used in this work). Additionally, to prevent stars close to the boundary from scattering the other sample due to uncertainties, stars within 0.02 dex of the boundary were removed from consideration entirely. The distributions in the \mgfe-\feh\ plane as well as the dividing line for both the APOGEE subgiants and APOKASC giants are shown in Figure \ref{fig:Alphapop}. The boundary we employ is given below:

\begin{equation}
    \begin{cases}
      \text{\mgfe} = 0.16, & \text{\feh} > 0.0 \\
      \text{\mgfe} = 0.16 - 0.13\text{\feh}, & \text{\feh} < 0.0\,~.
    \end{cases}       
\end{equation}

Stars with \mgfe\ above this line are classified as high-$\alpha$, whereas stars below are classified as low-$\alpha$. It is worth noting that although they are referred to as low-$\alpha$, they have [$\alpha$/Fe] ratios near that of the Sun.

\subsubsection{Pre-Dredge up Subgiants}\label{data:subgiants}

We use cool subgiants to measure the surface abundances prior to the FDU. Our subgiant sample was taken from APOGEE DR17. Asteroseismic data is not required since mass does not have an impact on the birth mixture. The FDU onset varies slightly with mass according to the MIST models, but according to APOGEE data shown in Figure \ref{fig:cnlogg}, it occurs between \logg\ of 3.5 and 3. To avoid contamination of mid-FDU stars, we consider only targets with \logg>3.5. Additional cuts in \logg\ and \teff\ space were placed to isolate the subgiants from the main sequence, and remove outliers from the main sample. These cuts were initially placed by visual inspection, but changing their position ``inward'' towards the main bulk of stars produced no change in our results, so they were deemed acceptable. The complete set of criteria are given below: 

\begin{equation}
    \begin{cases}
      \mathrm{\logg} \geq 3.5 \\
      \mathrm{\logg} \leq 0.004({T_\mathrm{eff}}) -15.7 \\
      \mathrm{\logg} \leq 0.0007 ({T_\mathrm{eff}}) + 0.36 \\
      \mathrm{\logg} \leq -0.0015 ({T_\mathrm{eff}}) +12.05 \\
      \mathrm{\logg} \geq 0.0012 ({T_\mathrm{eff}}) -2.8 \,~.
    \end{cases}       
\end{equation}

Figure \ref{fig:cnlogg} shows how our criteria yield a subgiant population (coloured in orange) this is cut off before significant drops in \cn\ take place. Also plotted are the MIST tracks, showing that these cuts are justified theoretically as well.

Finally, we limited the sample in \feh\ to focus on regions that have a sufficiently large population. The high- and low-$\alpha$ samples were each binned with a bin size of 0.02 dex. All stars in bins with fewer than 2\% of the maximum bin counts were removed from the sample. The boundaries for both populations are given below:

\begin{equation}
    \begin{cases}
      -0.75 \leq \mathrm{\feh} \leq 0.5 & \text{(Low-$\alpha$)} \\
      -1.15 \leq \mathrm{\feh} \leq 0.2 & \text{(High-$\alpha$)}\,~.
    \end{cases}       
\end{equation}

Our final sample includes 9372 low-$\alpha$ subgiants and 3517 high-$\alpha$ subgiants with reliable data.

\subsubsection{Lower Red Giant Branch}\label{data:LowRGB}

After the FDU, mass becomes an important parameter for understanding \cn, so the giant samples come exclusively from the APOKASC3 catalogue. Our first sample is made up of stars that have completed the FDU but have not yet begun to experience extra mixing effects. \changed{Extra mixing begins later on the RGB, around the luminosity of the RGB bump \citep{Gratton_2000}. The RGB bump is the point where stars stop expansion and re-contract for a short time, as the hydrogen-burning shell encounters the sharp chemical discontinuity left behind by the convective envelope.} We define the lower red giant branch stars (LRGB) by two criteria. First, the LRGB are stars that are flagged as RGB stars with the evolutionary state flag in APOKASC3. Second, we require that the LRGB stars have \logg\ values greater than 2.5, the \logg\ domain where extra mixing is not seen \citep{Shetrone_2019, Tayar&Joyce_2022}.

For low-$\alpha$ stars, we then applied quality cuts to remove outliers and stars with spurious values. First, we removed stars with a mass greater than 2 \msun. LRGB stars above this mass are extremely rare in our sample, so we did not compute the mass-\cn\ relationship beyond that mass. After that, we applied a weak \cn\ cut to remove stars that had exceptionally low \cn, clearly outside of the standard range of the sample. We believe these stars fall into the populations of chemically peculiar ``N-rich'' stars that have complicated origins \citep[e.g.][]{Johnson_2007, Fernandez-Trincado_2016, Martell_2016, Fernandez-Trincado_2022} and do not represent the typical star chemistry we are trying to define. Finally, similar to the subgiants, histograms were created on input parameters (\feh, \logg, \teff, and mass), and boundaries were set where bin populations dropped below 2\% of the densest bin value (rounded to nearer whole values for convenience). We believe that restricting discussion to regions well-populated by the sample provides higher accuracy analysis. We also enforced a maximum \logg\ of 3.25 to ensure that all stars used in the LRGB sample had completed FDU. The full criteria used for the low-$\alpha$ LRGB sample is detailed below:

\begin{equation}
    \begin{cases}
      \text{Evolutionary State: RGB} \\
      2.5 \leq \mathrm{\logg} \leq 3.25 \\
      4450 \leq T_\mathrm{eff} (\mathrm{K}) \leq 5050 \\
      -0.6 \leq \mathrm{\feh} \leq 0.45 \\
      0.8 \leq \mathrm{Mass (M}_\odot) \leq 2 \\
      \mathrm{\cn} \geq -0.6\,~.
    \end{cases}       
\end{equation}

For the high-$\alpha$ sample, the \logg, \teff, \cn, \feh, evolutionary state criteria were determined in the same manner as before. The mass cuts, however, needed to be determined differently.
Because the high-$\alpha$ population is predominantly old, nearly the entire sample is around the same mass: 1.08 \msun\ at \feh=0. There is a tail of higher mass, young stars in this sample, but these stars are likely the results of non-standard single-star evolution \citep{Jofre_2016, Jofre_2023}. We focus on the main population, for which there is a statistically significant sample, and remove the tail. Within the main population, more metal-rich stars have slightly higher masses on average. A mass cut was required to remove stars that deviate from this slight trend, having unusually high or low mass for stars at their \feh. The distribution of masses was treated as a Gaussian core with a tail towards the high mass and a metallicity term. The width of the Gaussian was set to be the median uncertainty in the mass, and stars more than 2 sigma from the mean mass at their \feh\ were removed. The final cuts employed on these stars are given below:

\begin{equation}
    \begin{cases}
      \text{Evolutionary State: RGB} \\
      2.5 \leq \mathrm{\logg} \leq 3.25 \\
      4450 \leq T_\mathrm{eff} (\mathrm{K}) \leq 5050 \\
      -0.9 \leq \mathrm{\feh} \leq 0.2 \\
      \mathrm{Mass (M}_\odot) \geq 1.01 + 0.15*\mathrm{\feh} \\ 
      \mathrm{Mass (M}_\odot) \leq 1.14 + 0.15*\mathrm{\feh} \\
      \mathrm{\cn} \geq -0.6\,~.
    \end{cases}       
\end{equation}

The LRGB stars are shown in blue in Figure \ref{fig:HR}. After all cuts, there are 2653 low-$\alpha$ and 434 high-$\alpha$ LRGB stars that we use in our analysis. The empirical boundaries we adopt for the completion of the FDU and the RGB Bump do not align perfectly with theoretical predictions. However, modest shifts in these boundaries do not meaningfully impact our fits found later in the paper. 

\subsubsection{Upper Red Giant Branch}\label{data:HighRGB}

Further along the RGB, extra mixing could become relevant, and so we define a new sample: the upper red giant branch stars (URGB). These are still first-ascent giants but have evolved to the point of becoming subject to possible extra mixing effects.

The criteria for these stars is much like the LRGB, requiring the evolutionary state flags to indicate that they are RGB stars, but now restricted to gravities below 2.5. This places them largely above the RGB bump. The \feh, \logg, and \teff\ boundaries for the low-$\alpha$ URGB were found in the same manner as for the LRGB, but the mass boundary is increased to 3.3 \msun\ as higher-mass stars are not as rare in the URGB as for the LRGB. The \cn\ cut was kept identical as well because these samples are very similar in the low-$\alpha$ region. The list of criteria used is below:

\begin{equation}
    \begin{cases}
      \text{Evolutionary State: RGB} \\
      1.4 \leq \mathrm{\logg} \leq 2.5 \\
      3900 \leq T_\mathrm{eff} (\mathrm{K}) \leq 5000 \\
      -0.65 \leq \mathrm{\feh} \leq 0.4 \\
      0.8 \leq \mathrm{Mass (M}_\odot) \leq 3.3 \\
      \mathrm{\cn} \geq -0.6\,~.
    \end{cases}
    .
\end{equation}

For the high-$\alpha$ stars, the cuts were determined identically as the LRGB, producing the following criteria:

\begin{equation}
    \begin{cases}
      \text{Evolutionary State: RGB} \\
      1.4 \leq \mathrm{\logg} \leq 2.5 \\
      3900 \leq T_\mathrm{eff} (\mathrm{K}) \leq 5000 \\
      -0.9 \leq \mathrm{\feh} \leq 0.1 \\
      \mathrm{Mass (M}_\odot) \geq 0.96 + 0.2*\mathrm{\feh} \\ 
      \mathrm{Mass (M}_\odot) \leq 1.18 + 0.2*\mathrm{\feh} \\
      \mathrm{\cn} \geq -0.6\,~.
    \end{cases}       
\end{equation}

The URGB stars are shown in green in Figure \ref{fig:HR}. After all aforementioned cuts, there are 1910 low-$\alpha$ and 291 high-$\alpha$ URGB stars that we use in our analysis. It is worth noting that these two samples have the least reliable data of all the samples. Seismology is less reliable for the more luminous giants \citep[][Pinsonneault et al., in prep]{Zinn_2019, Zinn_2022}. Additionally, there is the possibility of contamination by AGB stars in this sample. Most AGB stars lie between the RC and the tip of the RGB in \logg-\teff\ space and are difficult to distinguish from RGB stars even with asteroseismology \citep{Kallinger_2012, Christensen-Dalsgaard_2014}. These stars could have different surface chemistry than the first ascent giants at similar \logg\ because they have completed the extra mixing and mass loss that occurs throughout the upper RGB. While AGB contamination in this sample could increase the scatter, there are no reliable ways to remove them, so we will simply acknowledge the possible limitations of this sample and examine the impact they may have.

\subsubsection{Red Clump Stars}\label{data:clump}

Our final evolutionary state is the red clump (RC) stars. These are stars that have evolved through the entire RGB and begun helium fusion in their cores. Because these stars have already gone through the entire first ascent giant phase, they have undergone the maximum amount of extra-mixing and are a useful test case of the strength of these effects. RC stars also typically experienced notable mass loss at the tip of the RGB \citep{Origlia_2014}, which must be considered and accounted for when comparing with first ascent giants. Fortunately, asteroseismology can reliably separate them from RGB stars.

The criteria for this sample were determined through the same methods as the previous giant samples. We required that the star be a member of the RC according to the APOKASC evolutionary state flags. General population cuts were made on \feh, \logg, \teff, and mass for low-$\alpha$ stars and were determined through the same process as the previous samples. The \cn\ cut to remove N-rich stars was shifted, however, as the RC stars occupy lower \cn\ ranges than first ascent giants. The list of criteria is given below:

\begin{equation}
    \begin{cases}
     \text{Evolutionary State: RC} \\
      2.2 \leq \mathrm{\logg} \leq 3.0 \\
      4400 \leq T_\mathrm{eff} (\mathrm{K}) \leq 5050 \\
      -0.6 \leq \mathrm{\feh} \leq 0.4 \\
      0.8 \leq \mathrm{Mass (M}_\odot) \leq 3.3 \\
      \mathrm{\cn} \geq -0.7\,~.
    \end{cases}       
\end{equation}

For the high-$\alpha$ stars, the cuts were again determined in an identical manner as the first ascent giants and the criteria we obtained are given below:

\begin{equation}
    \begin{cases}
     \text{Evolutionary State: RC} \\
      2.15 \leq \mathrm{\logg} \leq 3.0 \\
      4400 \leq T_\mathrm{eff} (\mathrm{K}) \leq 5050 \\
      -0.75 \leq \mathrm{\feh} \leq 0.1 \\
      \mathrm{Mass (M}_\odot) \geq 0.90 + 0.25*\mathrm{\feh} \\ 
      \mathrm{Mass (M}_\odot) \leq 1.16 + 0.25*\mathrm{\feh} \\
      \mathrm{\cn} \geq -0.7\,~.
    \end{cases}       
\end{equation}

The RC stars are shown in red in Figure \ref{fig:HR}. After all aforementioned cuts, there are 4245 low-$\alpha$ and 399 high-$\alpha$ RC stars that we use in our analysis.

\section{Empirically Quantifying \cn\ Trends}\label{sec:fitting}

To understand how \cn\ changes with mass and \feh, we find polynomial functions that describe the average behaviour of the samples. Our fitting procedures below differ slightly between the subgiants and giants because the surface \cn\ of the subgiants does not depend on mass as it does for the giants. Since our objective is to observe the broad, smooth trends of the data, we fit polynomials to binned medians rather than the entire sample. 

For the subgiant samples, both the low-$\alpha$ and high-$\alpha$ stars were sorted by \feh\ and then binned into groups of 200. Stars that were more than 5 standard deviations from the mean within a bin were considered outliers and removed. A second-order polynomial in \feh:
\begin{equation}\label{eq:subfeh}
    \text{[X/Y]} = a_2(\text{[Fe/H]})^2 + a_1(\text{[Fe/H]}) + a_0\,~,
\end{equation}
or in \mgh:
\begin{equation}\label{eq:submgh}
    \text{[X/Y]} = b_2(\text{[Mg/H]})^2 + b_1(\text{[Mg/H]}) + b_0\,~,
\end{equation}
was fit to the binned medians, which were given equal weight. For each fit, the large number of data points provided very small uncertainties in the fit parameters, so the measurement uncertainties dominate the error of the fit. The median measurement uncertainties were propagated through the function and combined with the uncertainty of the output parameter (such as \cfe\ or \cn\ depending on the function). This provided a metallicity-dependent error value for the function, which is shown on the plots as a shaded region.

We binned the high-$\alpha$ giants with nearly the same method because these stars are still effectively mono-mass and variations in \cn\ due to mass are small. No polynomials were fit to the high-$\alpha$ giants, and due to the smaller sample sizes, bins are adjusted to be smaller and are given where relevant.

For the low-$\alpha$ giants, bins were taken in two steps. The data was first rank-ordered in \feh\ and divided into 15 equally sized cohorts. These cohorts were then rank-ordered in mass and divided into 15 equally sized bins, for a total of 225 bins from the whole sample. Then each bin was checked for outliers and the median was taken through the same process as for the subgiant bins. This process was used for each low-$\alpha$ giant sample (LRGB, URGB, and RC). 

Finally, a polynomial function of the form:
\begin{equation}\label{eq:rgb}
\begin{aligned}
    \text{[C/N]} = & c_5(M/M_\odot)^2 + c_4(M/M_\odot) + \\
    &c_3(\text{[Fe/H]})^2 + c_2(\text{[Fe/H]}) + \\
    &c_1(M/M_\odot)(\text{[Fe/H]}) + c_0^{}\,~,
\end{aligned}
\end{equation}
was fit with a least-squares regression to the binned data. Errors for the fits are found similarly to the subgiants but median measurement uncertainties on both the input parameters of \feh\ and mass were propagated through the function.

\subsection{Birth [C/N]}\label{fits:subg}

\begin{figure}
    \centering
    \includegraphics[width=.45\textwidth]{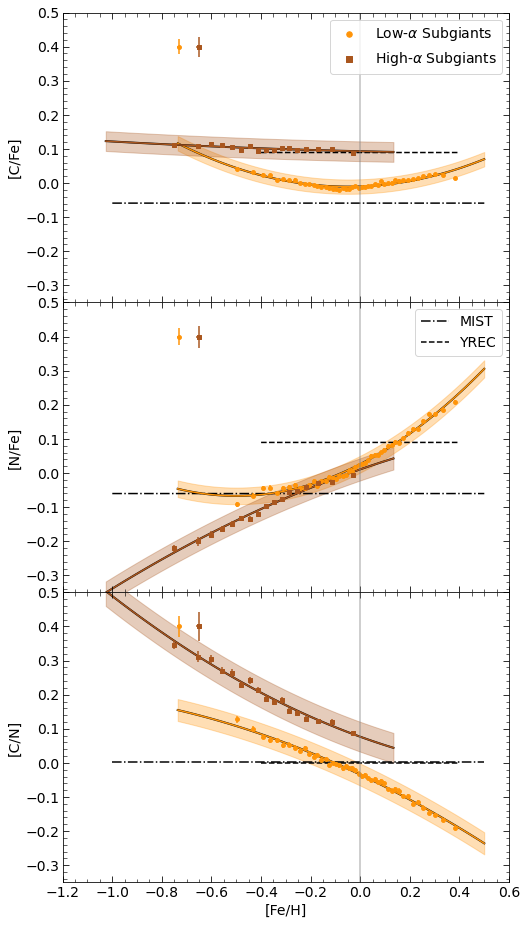}
    \caption{\cfe, \nfe, and \cn\ versus \feh\ for both low-$\alpha$ and high-$\alpha$ subgiant samples. The points represent bins of 200 stars with error bars representing the standard error of the values in the bin. The solid lines are the second-order polynomial fit of the data, and the shaded region indicates the error of the fit. The dashed and dash-dotted lines show the birth values employed by YREC and MIST models respectively. The single points in the corner of the plot show the median measurement errors of the respective parameters shown in the plot. The regression fit coefficients are given in Table \ref{tab:subgiantfe}}.
    \label{fig:cfebin}
\end{figure}

We see \changed{slight deviations from the solar mixture for carbon and strong} deviations for nitrogen in the pre-dredge-up subgiants. Figure \ref{fig:cfebin} shows that the high-$\alpha$ population has a higher \cfe\ at all \feh\ than the low-$\alpha$ population, but both have only slight slopes with \feh. Nitrogen, on the other hand, shows strong trends with \feh, with a similar slope in both $\alpha$-populations where they overlap in \feh. These two trends combine to create the observed behaviour of \cn, featuring both noticeable trends with \feh\ and offsets between $\alpha$ populations. To properly consider the birth abundance of \cn\ and the effects that it might have on post-FDU values, the $\alpha$ populations must be separated and \feh\ must be considered.

Theoretical evolutionary tracks, such as MIST or YREC \citep{yrec}, typically treat \cfe\ and \nfe\ abundances as constant, independent of \feh. This has been justified in part by a lack of data, which we have now provided. In our view, carbon and nitrogen should no longer be treated as solar-scaled in stellar models; instead, at minimum, they should be replaced with metallicity-dependent mass fractions relative to iron, and specific for high- and low-$\alpha$. This is likely to be important for both understanding the first dredge-up and low-temperature opacities, which \changed{are} sensitive to the CNO mixture.

Figure \ref{fig:cfebin} shows the birth abundances for MIST tracks and YREC models used in \citet{Tayar_2017}. Not only do these values differ from the empirical measurements, but they also disagree with each other. More discussion of the importance of birth trends on the final abundances will follow in Section \ref{sec:analysis}.

\begin{table}
    \centering
    \begin{tabular}{c c c c c}
         \multicolumn{2}{c}{Abundance} & $a_2$ & $a_1$ & $a_0$  \\
        \hline
         \multirow{3}{2.5em}{Low-$\alpha$} & [C/Fe] & 0.268 & 0.0258  & -0.00983 \\
         & [N/Fe] & 0.373 & 0.373 & 0.0260 \\
         & [C/N] & -0.118 & -0.344 & -0.0343 \\
         \hline
         \multirow{3}{2.5em}{High-$\alpha$} & [C/Fe] & -0.0121 & -0.0166 & 0.0933 \\
         & [N/Fe] & -.0983 & 0.251 & -0.0108 \\
         & [C/N] & 0.149 & -0.263 & 0.0767  \\
    \hline
    \end{tabular}
    \caption{Regression Coefficients for Subgiant Abundances in versus Fe}
    \label{tab:subgiantfe}
\end{table}

\begin{table}
    \centering
    \begin{tabular}{c c c c c}
         \multicolumn{2}{c}{Abundance} & $b_2$ & $b_1$ & $b_0$  \\
        \hline
         \multirow{3}{2.5em}{Low-$\alpha$} & [C/Mg] & 0.340 & 0.119 & -0.0923 \\
         & [N/Mg] & 0.430 & 0.475 & -0.0831 \\
         & [C/N] & -0.0860 & -0.351 & -0.00989 \\
         \hline
         \multirow{3}{2.5em}{High-$\alpha$} & [C/Mg] & 0.544 & 0.243 & -0.202 \\
         & [N/Mg] & 0.491 & 0.638 & -0.375 \\
         & [C/N] & 0.0385 & -0.402 & 0.171  \\
    \hline
    \end{tabular}
    \caption{Regression Coefficients for Subgiant Abundance in Mg}
    \label{tab:subgiantmg}
\end{table}

\begin{figure}
    \centering
    \includegraphics[width=.45\textwidth]{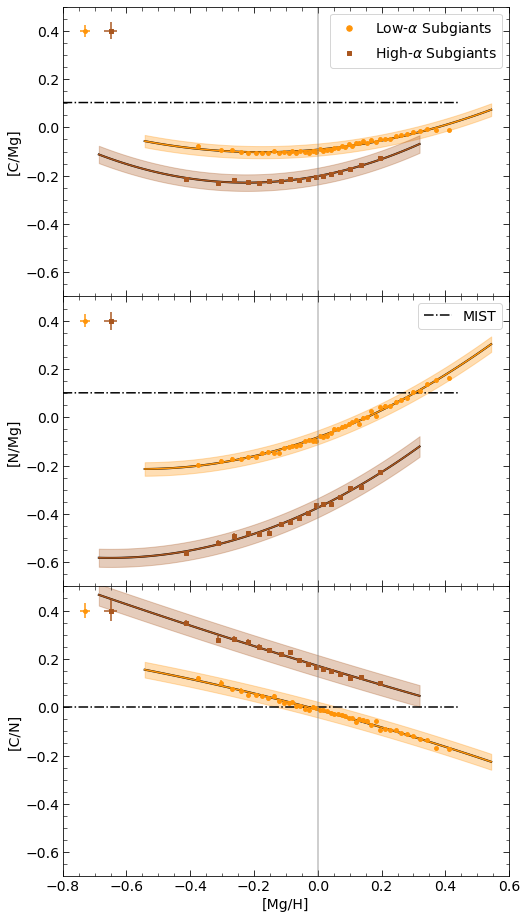}
    \caption{[C/Mg], [N/Mg], and \cn\ versus [Mg/H] for both the low-$\alpha$ and high-$\alpha$ subgiant samples. It has the same plotting conventions as Figure \ref{fig:cfebin}. The regression fit coefficients are given in Table \ref{tab:subgiantfe}}.
    \label{fig:cmgbin}
\end{figure}

While Fe is commonly used as a measure of overall chemical enrichment and as the reference element for abundance ratios, Mg is a physically simpler reference because it comes from a single, prompt nucleosynthetic source: core-collapse supernovae. For this reason, we also examined our subgiant samples in the planes of [C/Mg], [N/Mg], and \cn\ vs. \mgh, as shown in Figure \ref{fig:cmgbin}. [C/Mg] exhibits a small separation, about 0.1 dex, between the high-$\alpha$ and low-$\alpha$ populations and only weak metallicity trends within each population. [N/Mg] exhibits a much larger sequence separation, 0.29-dex at \mgh=0 and 0.34-dex at \mgh=-0.3. The metallicity trend of [N/Mg] is much stronger than that of [C/Mg], in both sequences. These [C/Mg] and [N/Mg] trends are qualitatively similar to those shown in Figure 9 of \cite{Vincenzo_2021}, which were derived by applying theoretical mixing corrections to APOGEE abundances of red giants using asteroseismic masses from \cite{Miglio_2021}. Our analysis of pre-FDU subgiants avoids the need for model-based ``unmixing'' corrections, and the consistency of results is reassuring.

The difference in [Fe/Mg] between the high-$\alpha$ and low-$\alpha$ sequences arises because time-delayed Type Ia supernovae have made a larger Fe contribution to the low-$\alpha$ stars. The degree of separation of [X/Mg] between these sequences is a diagnostic of the prompt vs. delayed nucleosynthesis contribution to element X \citep{Weinberg_2019, Weinberg_2022, Griffith_2019, Griffith_2022, Griffith_2023}. If we apply the methodology of Weinberg et al.\ (\citeyear{Weinberg_2022}, cf.\ their equations 6, 25, 26) to the [C/Mg] and [N/Mg] sequences shown in Figure \ref{fig:cmgbin}, we find that 72\% of C and 40\% of N in solar-abundance stars is produced by prompt sources, i.e., by core-collapse supernovae and massive star winds. For low-$\alpha$ stars with [Mg/H]=-0.3, the inferred fractions are nearly the same. We caution, however, that the \cite{Weinberg_2022} approach implicitly assumes that the delayed nucleosynthetic contribution tracks Fe from Type Ia supernovae, while for C and N the delayed contribution is presumably from AGB stars. For more accurate values one should construct chemical evolution models with realistic AGB delay times, as done by \cite{Johnson_2023} for N using the \cite{Vincenzo_2021} trends. We will pursue this approach using our empirical C trends in future work (D. Boyea et al., in preparation).

With these results in mind, the [C/N] vs. [Mg/H] trend in Figure \ref{fig:cmgbin} allows straightforward interpretation. Low-$\alpha$ stars have lower birth [C/N] than high-$\alpha$ stars because they have more time-delayed enrichment, and the fractional AGB contribution is larger for N than for C. [C/N] declines with increasing [Mg/H] for both populations because the N yield increases with metallicity, while the C yield is roughly independent of metallicity over the range of our sample. These conclusions are consistent with theoretical expectations. The interpretation of [C/N] vs.\ [Fe/H] in Figure \ref{fig:cfebin} is similar, though the detailed shape of the trends is different because of the mapping between [Mg/H] and [Fe/H] along the two sequences.

For further analysis of post-dredge-up stars, we return to using Fe as our reference, for consistency with prior published literature as well as being more widely available with more modest systematic offsets.

\subsection{\cn\ Immediately After First Dredge-up}\label{fits:lrgb}

\begin{figure*}
     \includegraphics[width=2.2\columnwidth]{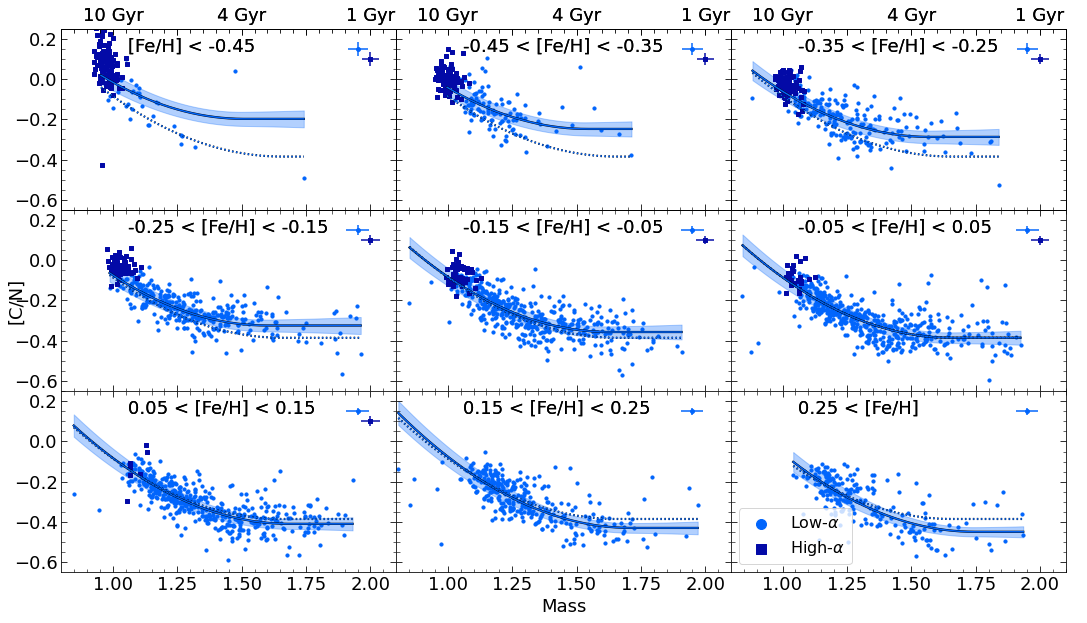}
    \caption{\cn\ versus mass for the LRGB samples, binned in \feh\. Corresponding MS lifetimes for solar metallicity stars are given at the top of the plot. The lighter circles show the low-$\alpha$ stars and the darker squares show the high-$\alpha$ stars. The fit shown was only made for the low-$\alpha$ stars, and the high-$\alpha$ are shown for comparison. The shaded region shows the uncertainty of the function. The dotted line represents the fit at solar metallicity and is shown in every window for reference. The function found for this data is found in Table \ref{tab:postdredge}.}
    \label{fig:lrgbpanel}
\end{figure*}

We now examine the FDU in isolation with the LRGB sample defined in Section \ref{data:LowRGB}. Figure \ref{fig:lrgbpanel} shows the best fit we obtain from the binned data points overlaid on the un-binned, low- and high-$\alpha$ LRGB samples. We find that the high-$\alpha$ stars fall slightly above the low-$\alpha$ stars in \cn\ at the same mass, but this offset is much smaller than observed in the pre-dredge-up subgiants.

For the mass-\cn\ relationship, we observe a steeper slope in \cn\ as a function of mass at low masses which flattens at high masses. To capture the curvature at low mass, we have adopted a quadratic fitting function. This function, however, produces un-physical upward curvature for higher masses. We therefore truncate the curve at the vertex of the parabola and treat the relationship at masses beyond this point as flat. Because we fit with second-order terms on both mass and \feh, the mass where this vertex occurs varies with metallicity, but for the LRGB sample, it occurs between a range of about 1.3-1.5 \msun. We also applied other functions to this data, such as higher order polynomials, functions with exponential decay terms, and functions with additional variables such as \logg\ or \teff. We found that these changes improved the fit by only negligible amounts at best, and in some cases produced worse overall performance. 

These qualitative trends match previous works, but with our larger samples, we see a stronger flattening effect than observed in previous works such as \cite{Martig_2016}. In addition to the flattening, the scatter of the sample also increases at higher masses. The standard deviation of the points about our fit hovers around 0.05 dex between 1 and 1.5 \msun, but is noticeably higher outside this range, increasing to over 0.15 dex at both edges of the mass distribution. The increased scatter, as well as the flattening of the relationship make \cn\ a poor mass proxy above 1.5 \msun\ for LRGB stars.

\begin{table*}
    \centering
    \begin{tabular}{c c c c c c c}
         \cn\ & $c_5$ & $c_4$ & $c_3$ & $c_2$ & $c_1$ & $c_0$  \\
        \hline
         LRGB & 0.698 & -2.311 & 0.267 & 0.404 & -0.406 & 1.528 \\
         URGB & 0.097 & -0.461 & 0.224 & 0.055 & 0.017  & 0.203 \\
         RC   & 0.164 & -0.806 & 0.186 & 0.499 & -0.385 & 0.494 \\
    \hline
    \end{tabular}
    \caption{Regression Coefficients for post-dredge-up, low-$\alpha$ samples}
    \label{tab:postdredge}
\end{table*}
    
\subsection{\cn\ For Later Stage Stars}\label{fits:urgbrc}

Next, we repeat the process for the LRGB on the URGB and RC samples. As mentioned previously, the URGB stars could show the first signs of extra mixing, and the RC stars should have a fully completed signature of extra mixing and mass loss.

\begin{figure*}
     \includegraphics[width=2.2\columnwidth]{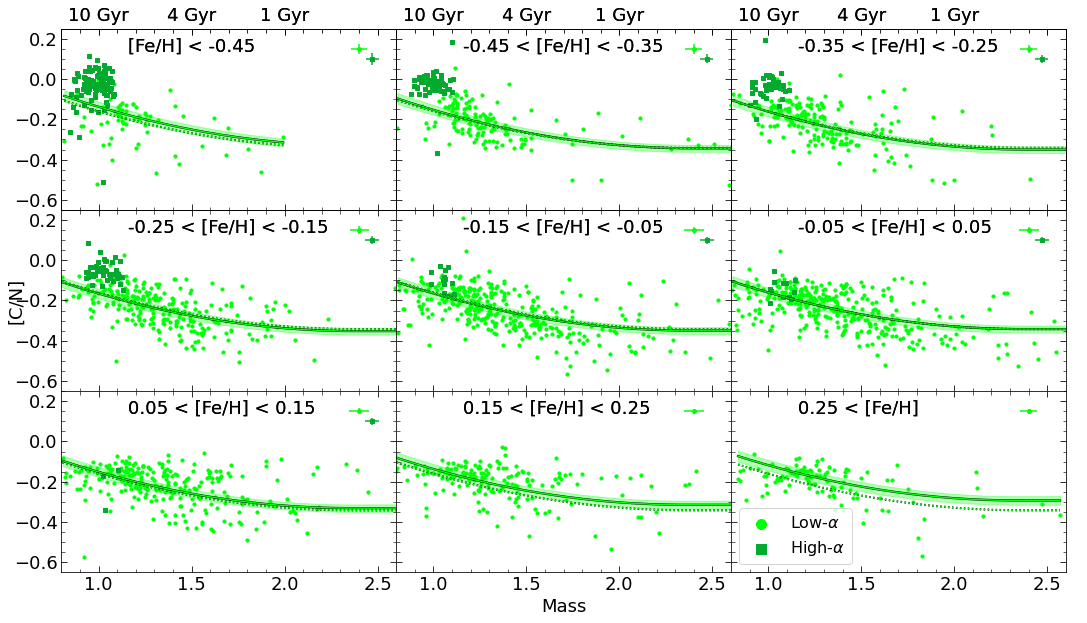}
    \caption{\cn\ versus mass for the URGB samples, in different \feh\ bins. The same conventions were used as for Figure \ref{fig:lrgbpanel}. The function found for this data is found in Table \ref{tab:postdredge}.}
    \label{fig:urgbpanel}
    \end{figure*}

Figure \ref{fig:urgbpanel} shows the URGB stars as well as the results of the fit. The high-$\alpha$ stars again show slightly enhanced \cn\ but not to the degree observed in the subgiants. However, the low-$\alpha$ stars here show significant deviations from the LRGB. First, there is a much weaker correlation between \cn\ and mass. The Spearman's rank correlation coefficient (Spearman's $\rho$) for the URGB sample is 0.59, as compared to 0.71 for the LRGB samples. Second, the scatter here also grows. There is no region of more extreme scatter, as with the LRGB sample, but the standard deviation of the residuals now fluctuates around 0.076. As previously stated in Section \ref{data:HighRGB}, these stars are expected to have higher errors and AGB contamination, so examining the RC will also help determine if this difference is due to physical processes, or observational limitations.

\begin{figure*}
     \includegraphics[width=2.2\columnwidth]{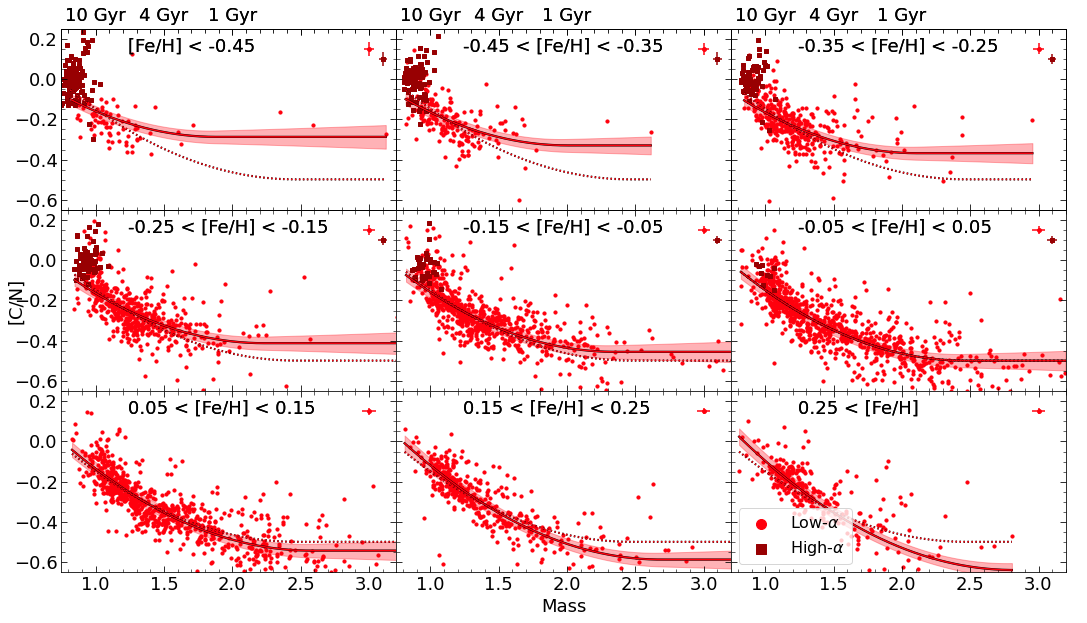}
    \caption{\cn\ versus mass for the RC samples, in different \feh\ bins. The same conventions were used as for Figure \ref{fig:lrgbpanel}. The function found for this data is found in Table \ref{tab:postdredge}.}
    \label{fig:rcpanel}
    \end{figure*}

The RC sample data and fits are shown in Figure \ref{fig:rcpanel}. Here, the trend again looks very much like the LRGB sample: strong correlations at lower masses, that gradually slacken as mass increases and \cn\ drops. The Spearman's $\rho$ for the RC is even higher than for the LRGB sample: 0.83. Additionally, while the shape of the curve is the same, the RC correlation extends into higher masses before flattening than the LRGB. The vertex and flattening of the curves in the RC sample occur between 1.7-2.5 \msun instead of the 1.3-1.5 \msun of the LRGB sample. This is a result of population effects. The weaker dependence of RC phase lifetime on mass results in much larger samples of high-mass stars measured with greater precision than LRGB stars. The LRGB and RC show very slight deviations in the regions where both are populated, so the extension of RC stars to higher masses is not an intrinsic difference in the \cn-mass relationship between the two populations, but rather a consequence of 3 \msun\ LRGB stars being too rare to adequately sample. Additionally, due to the similarities between the LRGB and RC samples, we can conclude the URGB sample's differences are the result of observational limitations on luminous giants, not the presence of extra mixing or mass loss, which is also present in the RC sample. 

\section{The First Dredge-Up, Extra Mixing, and Mass Loss}\label{sec:analysis}

With an understanding of the trends in \cn\ for stars before, during, and after RGB ascent, we quantify how much the mixing of the FDU changes the \cn\ on the surface of stars. The differences between these samples will reveal how much mixing takes place both during FDU, and after.

\subsection{Direct Impact of the First Dredge-Up}

\begin{figure}
    \centering
    \includegraphics[width=\columnwidth]{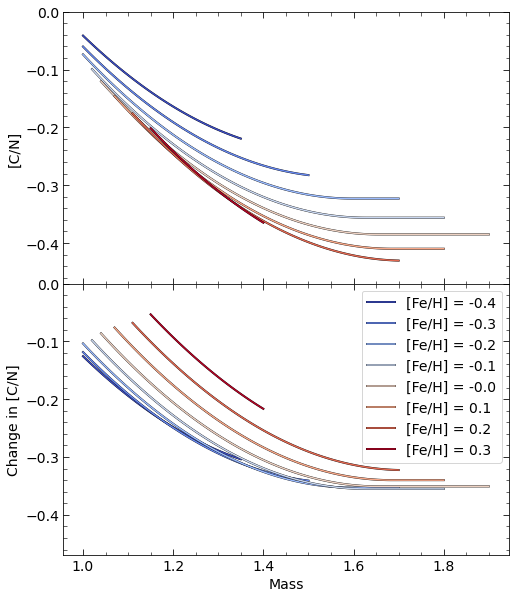}
    \caption{\changed{The difference between post-FDU abundances and pre-FDU abundances as a function of mass for the low-$\alpha$ samples. The top panel shows the difference from a flat, \cn=0 birth trend. The bottom panel shows the difference from the birth abundance trends found in section \ref{fits:subg}.} The \feh-mass space of the LRGB sample is not uniformly populated, primarily lacking high-mass, metal-poor stars. To reflect this, the range of the lines is restricted to match the space populated by the samples.}
    \label{fig:cndrop}
\end{figure}

\changed{By comparing the \cn\ of the subgiants of LRGB stars, we can directly examine how \cn\ changes from the FDU. Figure \ref{fig:cndrop} shows the change in \cn\ from the FDU for the low-$\alpha$ samples, according to the fits we found earlier. The top panel is the same function from section \ref{fits:lrgb}, reflecting the change in \cn\ from the birth mixture assuming a flat, \cn=0 trend, seen in MIST. The bottom panel shows the difference between the low-$\alpha$ subgiant abundances found in section \ref{fits:subg} and the LRGB abundances. This reflects the change in \cn\ from the observed birth abundance distribution.}

\begin{figure}
    \centering
    \includegraphics[width=\columnwidth]{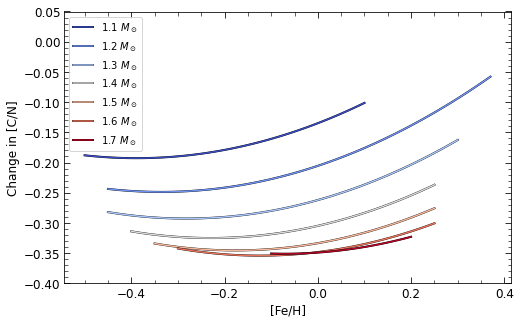}
    \caption{The same change in \cn\ function as shown in the bottom panel of Figure \ref{fig:cndrop}, but instead as a function of \feh\ at different mass values. The lines are limited to reflect the \feh\ range of the sample at that mass.}
    \label{fig:cndrop2}
\end{figure}

Though the mass dependence of the dredge-up is the same in both panels, the observed pattern with \feh\ changes entirely. In the top panel, higher \feh\ stars have less \cn\ after FDU, which seems to imply that these stars deplete their surface \cn\ more efficiently. However, after accounting for the birth abundances of these stars, we see that the inverse is true: stars with higher \feh\ show weaker depletion of surface \cn\ than other stars of equivalent mass. The observed decrease of surface \cn\ with \feh\ after the FDU is merely a consequence of these stars being born with less \cn. Additionally, this metallicity dependence weakens at higher masses, where the different trends begin to converge. This is highlighted even more strongly in Figure \ref{fig:cndrop2}, which shows the same trends accounting for birth abundances, but versus \feh. Not only does the \feh\ dependence weaken at high mass, but also at lower \feh, similarly to mass.

Figure \ref{fig:cndrop2} also shows which domains feature the most notable departure from their birth trends. Domains where these lines are flat, such as at high mass or low \feh, do not vary in the amount of \cn\ depleted from the surface. As such, in these domains, variations in the surface \cn\ after dredge-up are solely a result of variations in the birth \cn, and not due to differences in the amount of mixing they experience. 

\begin{figure}
    \includegraphics[width=\columnwidth]{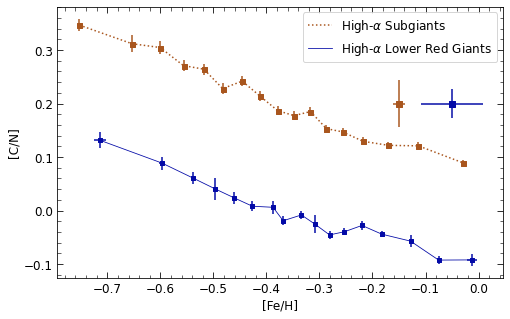}
    \caption{Pre- and post-dredge-up \cn\ values at various metallicities for the high-$\alpha$ sample. Here, we have elected just to use binned points to compare since both trends only depend on \feh\ and functions were not fit to high-$\alpha$ LRGB stars. The subgiants were put into bins of 200 as before, but due to the smaller size of the APOKASC3 sample, bins of 25 were used for the LRGB sample. Error bars shown on the points are the standard error of the distribution of the stars in the bin. The error bars on the points below the legend reflect the median individual measurement errors for these parameters of each sample.}
    \label{fig:highdredge}
\end{figure}

In the high-$\alpha$ domain, this is seen even more clearly. In Figure \ref{fig:highdredge}, the trends look essentially parallel, featuring $\sim$ 0.2 dex drop at all \feh. Unlike the trends seen in the low-$\alpha$ sample, the strength of the FDU in dex shows no variation with \feh. Even at \feh=0, where the low-$\alpha$ stars show variation in the dredge-up, the high-$\alpha$ stars maintain a constant change in \cn. Additionally, the degree of dredge-up is greater than expected for their mass. The high-$\alpha$ stars have a very small spread in mass, with a median of 1.02 \msun. The 0.2 dex drop observed in the high-$\alpha$ is $\sim$50\% greater than even the largest drop seen at this mass in the low-$\alpha$ sample. 

One potential complication to this interpretation would be age-metallicity relations. Because stars born more recently have higher metallicities, it is possible that the trends observed do not feature a flat dredge-up as we believe. If there are younger, higher-mass stars that experience different levels of dredge-up, they would live in the higher-metallicity range and have lower \cn\ because of a stronger FDU, rather than a difference in birth \cn. However, because the high-$\alpha$ stars not only have a small range in masses, but ages as well \citep{Miglio_2021}, we believe the impact of such an effect would be negligibly small. The dredge-up pattern may vary between low-$\alpha$ and high-$\alpha$ stars, but, the birth abundances have noticeable impacts on the post-dredge \cn\ values for both populations.

\subsection{Envelope Mixing Fraction}

\begin{figure}
    \includegraphics[width=\columnwidth]{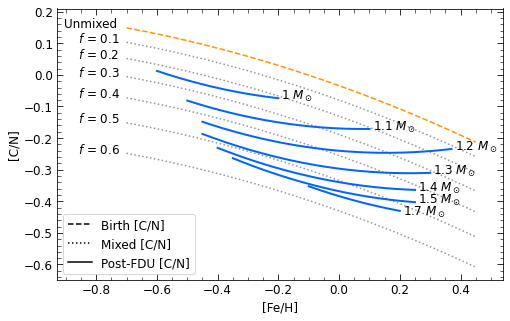}
    \caption{\cn\ versus \feh\ of the subgiants and the toy mixing model compared to the LRGB trends at various masses. The mixing fraction of each model is shown to the left of the line, The mass of each blue LRGB line is given to the right. Similarly to Figures \ref{fig:cndrop} and \ref{fig:cndrop2}, the range of the lines was limited to match the range populated by the samples. The intersection of the solid blue (observed) line with a dotted grey (model) curve indicates the fraction of the giants' convective envelopes that consists of CNO-processed material at the mass and \feh\ of the intersection.}
    \label{fig:toymodel}
\end{figure}

To translate the changes in \cn\ to an actual physical measure of mixing, we constructed a toy model to show how the surface abundance of a star is expected to change with different amounts of mixing. The model takes the birth \cn\ observed in the subgiants and dilutes it with some ``mixing fraction'' of CNO-processed material as a linear combination between the birth and processed \cn\ ratios. For example, a mixing fraction of 0.1 means a mixture that is comprised of 90\% birth mixture and 10\% CNO process material. The processed material is assumed to have a ratio of carbon to nitrogen atoms of 1/100. The true ratio is temperature dependent, but the difference is so great compared to surface abundances that changing the processed ratio by a factor of 2 in either direction has negligible effects on the abundances for fractions below 0.9 and no effects on our results. These fractionally mixed tracks were then compared with the fits of the post-FDU LRGB stars, at various masses, to see how the ``mixing fraction'' varies with mass and metallicity. Figure \ref{fig:toymodel} shows \cn\ versus \feh\ for the birth and mixed lines, as well as the post-FDU trends. 

The qualitative trends from Figures \ref{fig:cndrop} and \ref{fig:cndrop2} are once again apparent, with low-mass or metal-rich stars showing less mixing than high-mass or metal-poor stars. The degree of change is quite interesting, however. 1 \msun\ stars have roughly 25\% of their post-FDU convective envelope made up of CNO-processed material, whereas 1.7 \msun stars are all over 50\%. The impact of \feh\ is also apparent, as the 1.2 \msun stars vary between 10\% and $\approx45\%$ mixing fractions from metallicity alone, though the high and low-mass extremes do not show such sensitivity to metallicity. 

\changed{Figure 2 from \cite{Boothroyd&Sackmann_1999} shows how stellar models predict the deepest point of the dredge-up to vary with mass and metallicity. At a given metallicity, the deepest reaching mass coordinate decreases with mass until around 2.5 M$_{\odot}$. However, this scales in such a way that the unmixed mass is essentially constant. Metallicity has a comparatively small impact on the depth but does affect internal temperatures. Metal-rich stars are cooler, meaning their processing zones are smaller. At high mass, the amount of material is so large that the changes from metallicity are diluted, resulting in the higher mass stars showing less variation in mixing fraction with metallicity. At lower mass, there are stronger metallicity effects, but it is interesting to note how this combines with birth mixtures. Metal-rich stars pull less processed material to their surface, but also begin with a lower \cn, creating a sort of ``self-correction.'' The result is the flatter \cn\ curves are masses such as 1.2 \msun.}

\subsection{Effects from Extra Mixing and RGB Mass Loss}    

The FDU is not the only process that affects the \cn\ or mass of RGB stars. \changed{Although standard models predict constant surface abundances after the FDU, extra mixing is known to occur.} Changes to either the \cn\ or mass of a star after FDU can alter the relationship, and so comparing the relationships between different samples can illustrate the impact these effects have. \changed{Extra mixing has been observed in carbon isotope ratios and lithium abundances for many upper RGB stars \citep{Gilroy_1989, McCormick_2023}.} Additionally, while extra mixing has previously been observed in \cn\ in the high-$\alpha$ giants in APOGEE \citep{Shetrone_2019}, previous studies of low-$\alpha$ giants have not had such clear signals \citep{Souto_2019}. Additionally, studies such as \cite{Miglio_2012} and \cite{Tailo_2022} have examined mass loss on the RGB in open clusters. Using our data sets, we can follow up these studies by taking advantage of the newer, larger data sets to provide additional constraints on the strength of these effects, as well as in what regimes they apply.

\subsubsection{Extra mixing in High-$\alpha$ Stars}

\begin{figure}
     \includegraphics[width=\columnwidth]{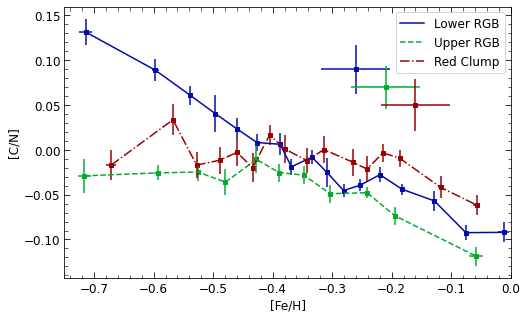}
    \caption{\cn\ versus \feh\ for high-$\alpha$ APOKASC giants. Each point represents a bin of 25 stars, with error bars representing their standard error. The error bars below the legend represent the median measurement errors of the parameters for each sample.}
    \label{fig:HighMixing}
    \end{figure}

The high-$\alpha$ stars present an easier sample to examine as these stars are all old and their relative mass is therefore similar. This means any RC star in the sample is a descendant of the RGB stars we observe. Mass loss does occur here, as the RC stars do have a lower median mass than the LRGB stars, but is not a concern since these stars are close to direct descendants. Figure \ref{fig:HighMixing} shows \cn\ versus \feh\ for all three giant samples. Below \feh=$-$0.4, the \cn\ of the URGB and RC deviate from the LRGB. Both the onset point and the dependence on \feh\ of extra mixing match the results from \cite{Shetrone_2019}. The magnitudes they report in Table 2 also largely match what we see, though we see slightly less mixing in the \feh=-0.4 region.

\changed{Above \feh=-0.4, an interesting picture arises. The RC stars have a higher \cn\ at the same metallicity than the LRGB stars. According to \cite{Shetrone_2019}, extra mixing does not appear to operate in the high-$\alpha$ stars in this domain,} which implies the three samples should lie on top of each other. However, the URGB is consistently lower than the LRGB, and the RC is consistently higher than the LRGB. Because RC stars are hotter on average than LRGB stars, and URGB are cooler, we believe this can be explained by a slight temperature systematic offset. We do note, however, that these effects are on the order of 0.03 dex, which is a plausible systematic error level, and smaller than the observational statistical errors. 

\subsubsection{Extra Mixing and Mass Loss in Low-$\alpha$ Stars}

Examining the low-$\alpha$ giants requires a slightly different process, as now mass becomes relevant again. It is important to note how extra mixing and mass loss will affect the trends we see. Extra mixing operates at low metallicities and lowers the surface \cn. Mass loss operates at low masses and lowers the mass. Because \cn\ and mass are negatively correlated, lowering the \cn\ or mass of the stars will appear identical, and so the main way of distinguishing these effects will be when they occur.

\begin{figure*}
     \includegraphics[width=2\columnwidth]{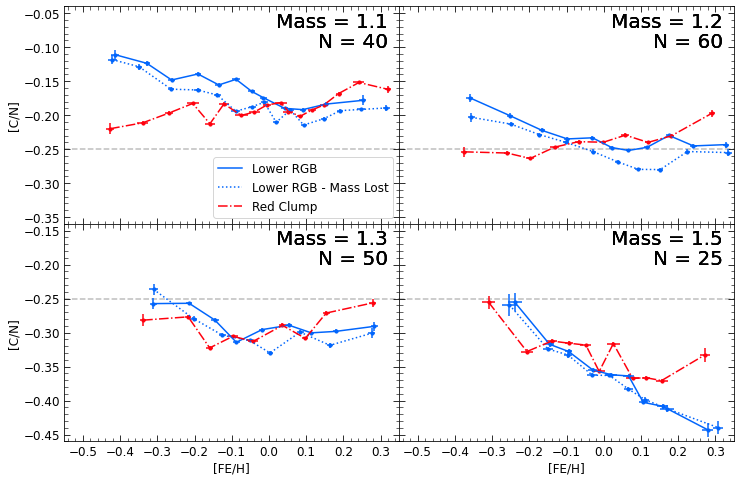}
    \caption{\cn\ versus \feh\ for low-$\alpha$ giants of various masses. The windows are 0.05 \msun in both directions of the listed mass. The number of stars in each bin vary between panels to have similar amounts of points between windows, and is listed as n in each panel. The y-axis limits where chosen to mirror the range spanned in Figure \ref{fig:HighMixing}. A dashed gray line is plotted at \cn=-0.25 in all plots to help show the decreasing \cn\ between mass bins.}
    \label{fig:LowMixing}
    \end{figure*}

\begin{figure*}
     \includegraphics[width=2\columnwidth]{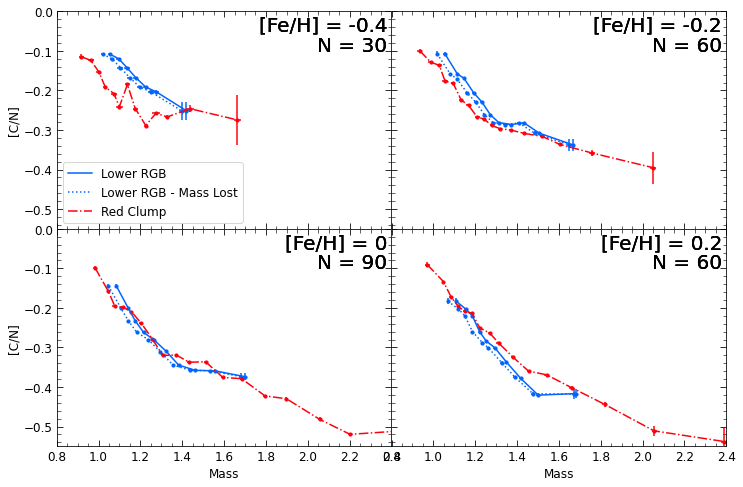}
    \caption{\cn\ versus mass for low-$\alpha$ giants of various \feh. All conventions in Figure \ref{fig:LowMixing} are kept here where relevant.}
    \label{fig:LowMixing2}
    \end{figure*}

To isolate the effects, we use MIST tracks to determine the expected degree of mass loss throughout the RGB evolution. We apply this mass loss function to our LRGB sample and compare that with the RC sample. Figures \ref{fig:LowMixing} and \ref{fig:LowMixing2} show bins of low-$\alpha$ stars in windows of mass and \feh\ respectively. The mass loss predicted by the MIST evolutionary tracks ranges between 0.005-0.040 \msun and produces subtle effects on the trends. The alterations to an observed \cn\ at a given mass due to mass loss are on the level of 0.02 dex. To concretely determine the presence of mass loss, spectroscopic abundances with uncertainties below that level would be required, which exceeds the current precision of the APOGEE abundances. Ultimately, mass loss produces a weak effect on the \cn-mass relationship.

Similarly, looking into the effects of extra mixing, we do not see compelling evidence in the low-$\alpha$ stars. Though the metal-poor RC stars do feature lower \cn\ than the metal-poor LRGB stars across most masses, the trend is reversed in the metal-rich end. For RC stars to have higher \cn\ than an LRGB of the same metallicity and mass would require ``un-mixing'' which is non-physical. Because the trends on one end of the domain mirror the other and are of similar scale, we do not find compelling evidence of extra-mixing in \cn\ of the low-$\alpha$ giants.

\changed{This may at first seem strange as extra mixing in carbon isotopic ratios and lithium has been observed in these same domains \citep{Gilroy_1989, McCormick_2023}. The lack of clear mixing signature in \cn\ for these stars does not imply there is no mixing at all, but rather it does extend deep enough to reach the nitrogen-enhanced material. The regions of nitrogen-enhanced material lie below the C$^{13}$ enhanced and lithium-destroying regions, where the temperature is higher \citep{Pinsonneault_1989}. Likely, these extra mixing effects in the low-$\alpha$ stars do not penetrate deep enough into the burning region to alter surface \cn\ ratios, while still being able to affect the carbon isotopic ratios and deplete surface lithium.}

\section{Discussion}\label{sec:discussion}
\subsection{Limits on \cn\ as a Mass Indicator}\label{dis:limits}

The \cn-mass relationship detailed above is useful, however, we must note its limitations as a mass diagnostic. The relationship weakens at higher masses and cannot be properly calibrated in certain regimes due to the rarity of such stars. For practical use in estimating masses, a robust determination of where this relationship can be applied is required.

\begin{figure*}
    \includegraphics[width=2.1\columnwidth]{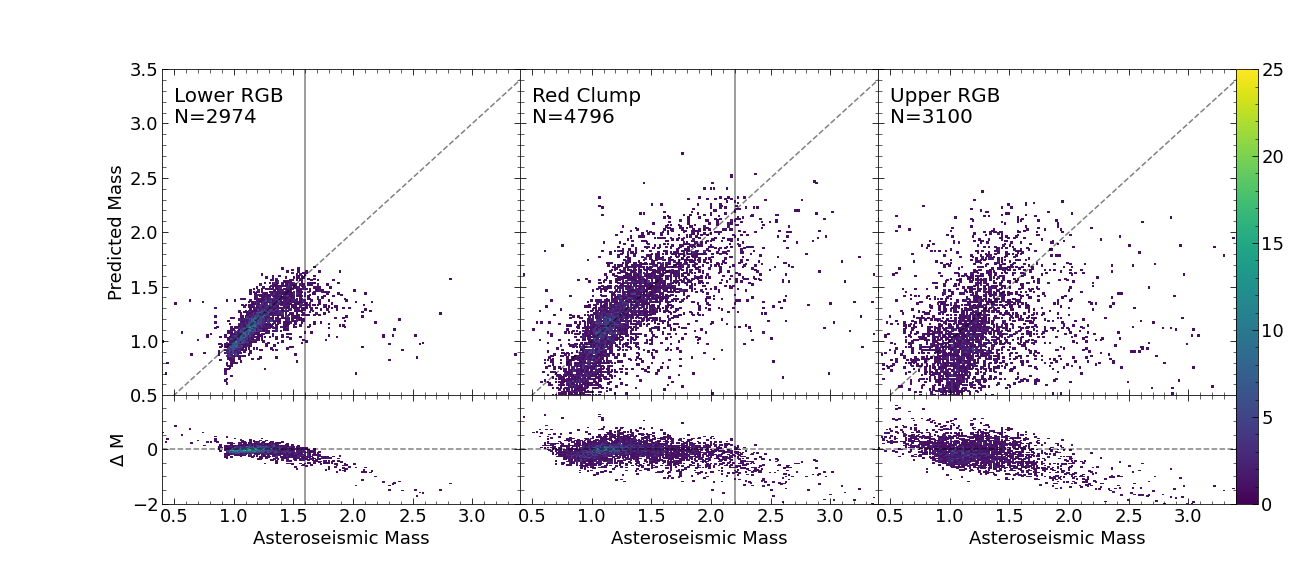}
    \caption{Mass estimated from \cn\ versus asteroseismic mass for full the APOKASC3 sample, separated by evolutionary state. The lower panel shows the absolute difference between the masses versus asteroseismic mass. The same bins have been used between all three panels. Vertical grey lines are drawn where the masses become under-predicted for the LRGB and RC samples.}
    \label{fig:massmass}
\end{figure*}

We performed a recovery test, recording the differences between the mass expected from \cn\ and the asteroseismic mass. The mass estimate was found by solving the quadratic equation for mass and adopting the lower solution where it is double-valued. At each metallicity, there is a domain in \cn\ for which there is no corresponding mass (\cn$<-0.4$ for \feh$=0$). Stars in this range were treated as invalid. Figure \ref{fig:massmass} shows the estimated mass versus the asteroseismic mass for the entire low-$\alpha$ APOKASC3 sample, separated by evolutionary state.

The fits perform well in regions that are well populated, but begin under-predicting mass when we extend into the regime where the fit was flattened. The LRGB sample shows a tighter correlation, with less scatter than the other samples, but begins under-predicting masses around 1.6 \msun. The RC sample shows a greater scatter overall but does not begin under-predicting masses until above 2-2.2 \msun. The URGB sample performs sub-optimally. As discussed previously, the higher errors on the measurements for these stars make this an expected effect. Likely, many of the stars with asteroseismic masses above 2 \msun\ in this sample are not truly high-mass, but rather lower-mass stars scattered up into that domain. Overall, for the less luminous giants, the \cn-mass relationship recovers the asteroseismic mass well. 

However, Figure \ref{fig:massmass} does not tell the whole story. Stars in some domains are better fit by the relationship than others, as shown in Figure \ref{fig:recovery}. Because some stars are estimated poorly, while others are beyond the domain of the function, we consolidate these two cases into what we call failures. If a star had a predicted mass that differed from the catalogue mass by 20\% or more or lay within a region where we were unable to get the predicted mass, it was labelled a failure. LRGB stars recovery is largely a function of mass: low-mass stars perform well across all metallicities, and high-mass stars perform poorly. The RC sample is slightly different. Low mass, metal-poor stars are not recovered consistently, and the effective mass range grows at higher \feh. The LRGB and RC samples are complementary in that regard. The LRGB fits low mass stars more reliably across the full metallicity range we observe, whereas the RC can probe higher masses than the LRGB can. 

These domains of effectiveness are even sharper in observational space, shown in Figure \ref{fig:recovery2}. In the \cn-\feh\ plane, nearly all the failures in the LRGB sample are stars with \cn\ outside of the domain of the relationship, shown in the dark red zone with a sharp boundary on the bottom of the top left panel. For the RC stars, a similar boundary appears, but the relationship also fails for the high \cn\ stars. However, in both cases, there are clear regions in the observed space where this relationship can be reliably applied. The URGB, unsurprisingly, does not have any region where it reliably exceeds a 50\% success rate on recovering asteroseismic masses. 

For the LRGB and RC samples, we provide boundaries in the \cn-\feh\ plane that define the region where the relationship performs reliably as a mass diagnostic. They are given below:

\begin{figure}
    \includegraphics[width=\columnwidth]{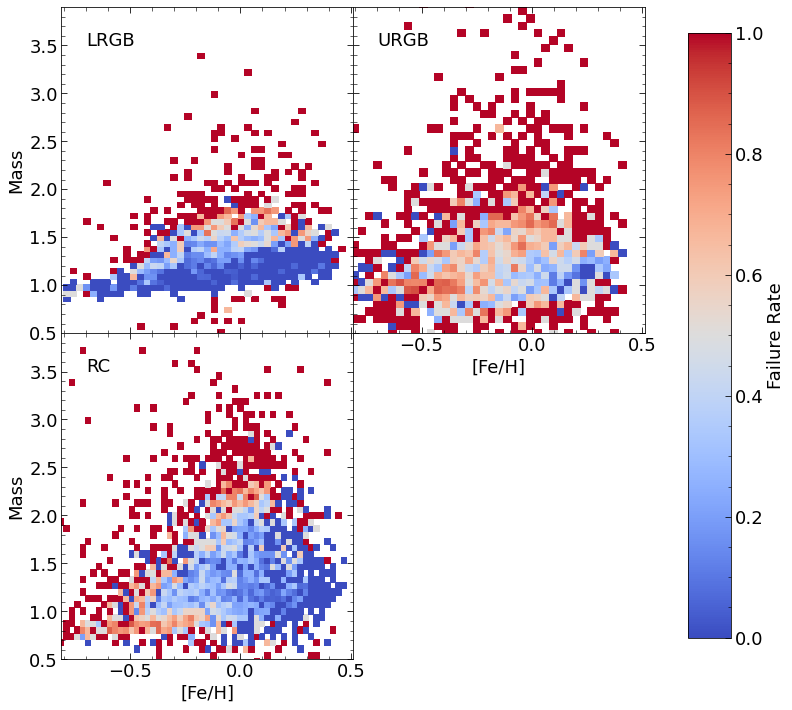}
    \caption{Histogram of the recovery test results in mass-\feh\ space. Each bin is coloured by the fraction of stars in the bin that are considered failed recoveries (no mass or mass error > 20\%). All APOKASC stars not removed for complicating phenomena or incomplete entries are shown here.}   
    \label{fig:recovery}
\end{figure}

\begin{figure}
    \includegraphics[width=\columnwidth]{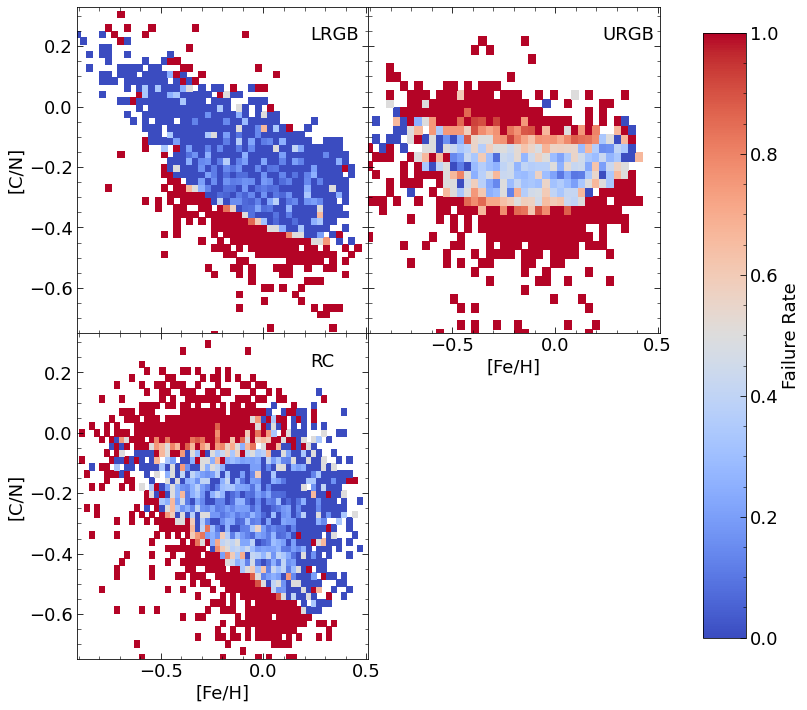}
    \caption{Same as Figure \ref{fig:recovery}, but in \cn-\feh\ space.}
    \label{fig:recovery2}
\end{figure}

\begin{equation}
    \begin{cases}
     \text{LRGB} \\
      -0.8 \leq \mathrm{\feh} \leq 0.4 \\
      \mathrm{\cn} > (0.208*\mathrm{\feh}^2) + (-0.268*\mathrm{\feh}) + (-0.385)  \\
      \mathrm{\cn} < (-0.1*\mathrm{\feh}) + (-0.06)  \\ 
    \end{cases} 
    \label{eq:LRGBvalid}
\end{equation}

\begin{equation}
    \begin{cases}
     \text{RC} \\
      \mathrm{\feh} \leq 0.4 \\
      \mathrm{\cn} > (-0.5*\mathrm{\feh}) + (-0.45)  \\
      \mathrm{\cn} < -.1  \\ 
    \end{cases}  
    \label{eq:RCvalid}
\end{equation}

Ultimately, \cn\ has promise as a mass indicator, but not universally. The optimal use cases also vary with the stage of the star you examine. To examine lower-mass stars, LRGB behave the best as the relationship holds across a large metallicity range and sports the lowest errors. The RC does allow mass estimates into a higher-mass domain, though it is limited at lower metallicities and has a higher scatter. 

\subsection{Offsets between APOGEE Data Releases}

Our analysis in this paper relies on data from the APOGEE-2 DR17. However, offsets between different data releases exist because of changes to the analysis pipelines. We take the shifts between two data releases as an indication of the scale of difference that can be expected from observational errors. By comparing DR16 and DR17 abundances, and applying our fits to the DR16 abundances, we can examine how much the mass estimates are affected by the choices made during abundance analysis.

Examining stars that are in both catalogues and were used in our analysis, we find the DR16 \cn\ values to be roughly 0.06 dex higher on average than the DR17 values. When applying our fits trained on DR17 to the DR16 data, we find that stars' predicted mass decreases by about 0.1-0.25 \msun, with larger differences in RC stars than LRGB stars. Of course, if we had calibrated our formulae to DR16 \cn\ values, then their application to DR16 abundances would have given masses closer to what we found for the same stars in DR17. However, it is worth considering the effects these can produce, as we are limited by the accuracy of the APOGEE abundance scale.

\subsection{Comparison with Previous Work}

\begin{figure*}
    \includegraphics[width=2.1\columnwidth]{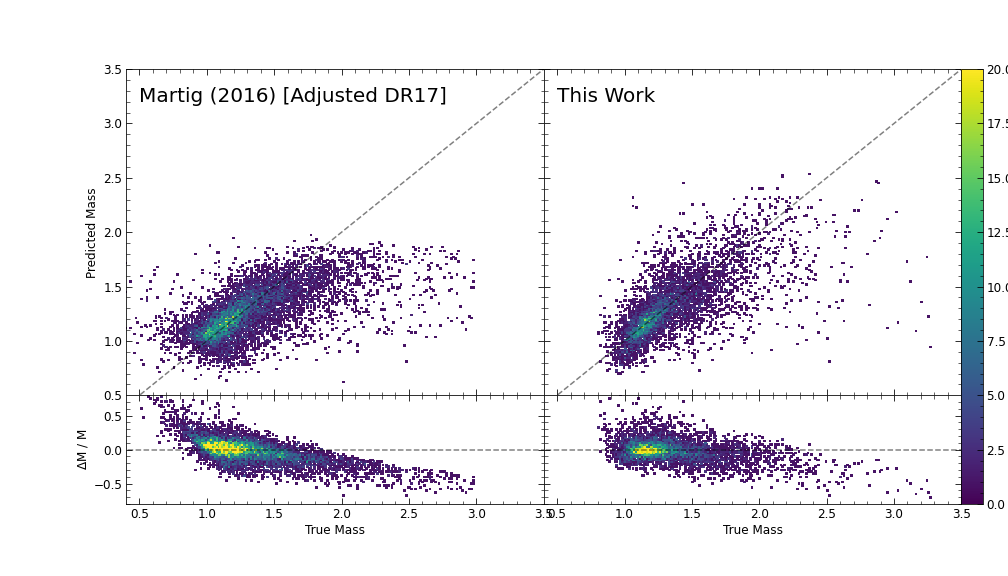}
    \caption{Predicted mass versus asteroseismic mass for both this work and \protect\cite{Martig_2016} (Table A1). The bottom panels show relative difference in mass to mimic Figure 7 from their paper.}
    \label{fig:martigcompar}
\end{figure*}

\cite{Martig_2016} obtained mass and age relationships between carbon, nitrogen, and metallicity using APOGEE DR12 abundances and APOKASC2 masses. To directly compare our results with theirs, we first take the sample of stars in both catalogues to find the relationship between DR12 and DR17 abundances to scale between the two. We then apply the scaling to all stars within our sample and apply their fit (from Table A1) to our stars that fall within their specified effective region and obtain predicted masses for our sample. We use their Table A1 fit as opposed to their Table A2 fit due to difficulty in implementation. Even after scaling quantities back to DR12 values, there were still a large number of erroneous, negative mass values produced. This is likely due to various changes in the APOGEE pipelines that have occurred, but we were unable to appropriately apply the more comprehensive fit to our data. In Figure \ref{fig:martigcompar}, we plot the density of points in true mass versus predicted mass space for both their fit and our fit. For our fit, we adopt the boundaries given in section \ref{dis:limits}. 

\section{Summary and Conclusions}\label{sec:conclusion}

Using the APOGEE DR17 and APOKASC3 datasets, we define samples of subgiants and giants separated by $\alpha$-enrichment and evolutionary stage. We analyze the evolution of \cn\ from the subgiants through the lower red giant branch, upper red giant branch, and red clump. Through contrasting the \cn\ at different stages we look for the impact of chemical evolution, first dredge-up, extra mixing, and mass loss on the \cn-mass relationship and have the following findings:

\begin{itemize}
    \item The birth \cn\ of stars varies with \feh\ and $\alpha$-enrichment. High-$\alpha$ stars show enhanced \cn\ values relative to low-$\alpha$ stars of similar \feh, and both populations show strong negative correlations between birth \cn\ and \feh. These trends are a consequence of a large delayed contribution to N enrichment that increases with \feh. Solar-scaled values are a poor approximation of the data. \\
    
    \item The post-FDU \cn\ ratios of giants are impacted by variations in their birth \cn. The observed \feh\ dependence of the dredge-up is a combination of differences of mixing as a function of \feh, as well as differences in birth \cn. The effects of the birth \cn\ are also partially self-corrected by the difference in mixing strength as a function of \feh. \\
    
    \item \cn\ is an effective mass diagnostic for first ascent giant stars below the RGB bump up to around 1.5 \msun. Beyond this point, the \cn\ differences between masses become small and difficult to separate.\\
    
    \item Red clump stars also support a strong \cn-mass relationship, and extend the application of \cn\ as a mass indicator to higher masses, but the range of application varies strongly with \feh. \\
    
    \item We do not detect a signal of RGB mass loss in the \cn\ of low-$\alpha$ stars. \\
    
    \item Unlike lithium and carbon-isotope ratios, \cn\ does not show compelling evidence of extra-mixing in low-$\alpha$ giants. \\
    
    \item We provide theoretical estimates of fractions of processed material in post-FDU giants for testing stellar interior models.
    
\end{itemize}

\cn\ as a mass diagnostic has promising power, but does have some limits. The \cn-mass relationship fails in certain regimes, though using both RGB and RC stars together can extend the regimes over which the relationship can be applied. Additionally, we are ultimately limited by the precision of the APOGEE abundance scale, which can alter mass predictions by a small, but non-negligible amount. 

Future surveys, such as Milky Way Mapper, will provide a much larger sample of stars with high-quality spectroscopic abundances. With the addition of asteroseismic parameters from surveys such as TESS \citep{TESS}, K2 \citep{K2}, and PLATO \citep{Plato}, even larger samples of stars with known masses and surface chemistry will be available, which will not only allow such relationships to be extended into larger regions of parameter space but also be able to provide a large enough sample of stars to find possible causes of deviations from these relationships.

\section*{Acknowledgements}

We thank the referee for their helpful comments.

AS is partially supported by MICINN grant PID2019-108709GB-I00, the program Unidad de Excelencia Maria de Maeztu CEX2020-001058-M, and EU program ChETEC-INFRA under grant agreement 101008324.
R.A.G. acknowledges the support from PLATO CNES grant. M.V acknowledges support from NASA grant 80NSSC18K1582.
S.M.~acknowledges support from the Spanish Ministry of Science and Innovation (MICINN) with the Ram\'on y Cajal fellowship no.~RYC-2015-17697, PID2019-107187GB-I00, the grant no. PID2019-107061GB-C66, and through AEI under the Severo Ochoa Centres of Excellence Programme 2020--2023 (CEX2019-000920-S). S.H. acknowledges support from the European Research Council via the ERC consolidator grant 'DipolarSound' (grant agreement \# 101000296).

This work made use of Astropy:\footnote{http://www.astropy.org} a community-developed core Python package and an ecosystem of tools and resources for astronomy \citep{astropy2013, astropy2018, astropy2022}.

Funding for the Sloan Digital Sky Survey IV has been provided by the Alfred P. Sloan Foundation, the U.S. Department of Energy Office of Science, and the Participating Institutions. 

SDSS-IV acknowledges support and resources from the Center for High Performance Computing  at the University of Utah. The SDSS website is www.sdss4.org.

SDSS-IV is managed by the Astrophysical Research Consortium for the Participating Institutions of the SDSS Collaboration including the Brazilian Participation Group, the Carnegie Institution for Science, Carnegie Mellon University, Center for Astrophysics | Harvard \& Smithsonian, the Chilean Participation Group, the French Participation Group, Instituto de Astrof\'isica de Canarias, The Johns Hopkins University, Kavli Institute for the Physics and Mathematics of the Universe (IPMU) / University of Tokyo, the Korean Participation Group, Lawrence Berkeley National Laboratory, Leibniz Institut f\"ur Astrophysik Potsdam (AIP),  Max-Planck-Institut f\"ur Astronomie (MPIA Heidelberg), Max-Planck-Institut f\"ur Astrophysik (MPA Garching), Max-Planck-Institut f\"ur Extraterrestrische Physik (MPE), National Astronomical Observatories of China, New Mexico State University, New York University, University of Notre Dame, Observat\'ario Nacional / MCTI, The Ohio State University, Pennsylvania State University, Shanghai Astronomical Observatory, United Kingdom Participation Group, Universidad Nacional Aut\'onoma de M\'exico, University of Arizona, University of Colorado Boulder, University of Oxford, University of Portsmouth, University of Utah, University of Virginia, University of Washington, University of Wisconsin, Vanderbilt University, and Yale University.

\section*{Data Availability}

The SDSS-IV APOGEE Data Release 16 and 17 is available at https://www.sdss4.org. The APOKASC3 data used for this paper are available on request and will be published shortly. The MIST isochrones
are available at https://waps.cfa.harvard.edu/MIST/.



\bibliographystyle{mnras}
\bibliography{biblio}




\bsp	
\label{lastpage}
\end{document}